\documentclass[10pt]{article}
\usepackage[utf8]{inputenc}
\usepackage{amsmath}
\usepackage{amsfonts}
\usepackage{amssymb}
\usepackage{extsizes}
\usepackage{graphicx}
\usepackage{yfonts}
\usepackage{tikz}
\usepackage[left=2cm,right=2cm,top=2cm,bottom=2cm]{geometry}

\title{\textbf{Fingerprints of the quantum space-time in time-dependent quantum mechanics: An emergent geometric phase}}
\author{
	{\bf {\normalsize Anwesha Chakraborty}
		\thanks{anwesha@bose.res.in }},
	{\bf{\normalsize Partha Nandi}
	\thanks{parthanandi@bose.res.in}},
	{\bf {\normalsize Biswajit Chakraborty}
		\thanks{dhrubashillong@gmail.com}}
	\\
 {\normalsize Department of Theoretical Sciences}\\
	{\normalsize S.N. Bose National Centre for Basic Sciences}\\
	{\normalsize JD Block, Sector III, Salt Lake, Kolkata 700106, India}\\}
\begin{document}
\maketitle
\begin{abstract}
We show the emergence of Berry phase in a forced harmonic oscillator system placed in the quantum space-time of Moyal type, where the time $\textquoteleft{t}$' is also an operator. An effective commutative description of the system gives a time dependent generalised harmonic oscillator system with perturbation linear in position and momentum. The system is then diagonalised to get a generalised harmonic oscillator and then its adiabatic evolution over time-period $\mathcal{T}$ is studied in Heisenberg picture to compute the expression of geometric phase-shift. 
\end{abstract}
\section{Introduction}
The general consensus is that the quantum gravitational effects are expected to be observed only near the Planck length scale $l_p=\sqrt{\frac{\hbar G}{c^3}} \approx 10^{-33} $ cm. It was first proposed by M. P. Bronstein \cite{bron} and thereafter Doplicher et. al. \cite{dop} that, at very low length scale, like in the vicinity of Planck length scale, the concept of localization of an event in space-time breaks down in the sense that any attempt towards this localization will inevitably result in gravitational collapse. One way to avoid such a scenario is to
postulate a non-commutative (NC) algebra between the space-time coordinates, which are now promoted to the level of operators. It thus becomes quite fruitful and interesting to build models of quantum mechanics (QM) and quantum field theory (QFT)
in such length scale. However, the  connection between quantum gravity and quantum field theory  still remains quite elusive, primarily due to this localisation problem. So therefore it becomes quite imperative to study different aspects of quantum space-time including the case where the time is also an operator and for that one clearly needs to take the first step towards the formulation of a consistent version of QM in quantum space-time and eventually the QFT. In this context, we can point out some earlier works in this direction \cite{szabo,apb,qft,liz} where some fascinating results were obtained, like discretization of time \cite{liz,time} corroborating similar observations made earlier by 't Hooft in the context of (2+1) dimensional quantum gravity \cite{hooft} and correction of spin-statistics theorem in NC spacetime leading to non-Pauli like transitions \cite{bal2}. Although, the effects of non-commutativity  in the space-time sector should presumably become significant at (or before) very high energy scale, for instance, close to Planck scale energy, it is intriguing to speculate that there should be some relics of the effects of spacetime non commutativity in the low energy regime\cite{sg1}, because of the inadequate decoupling mechanism between the high and low energy sectors. In fact in \cite{partha}, some of us have shown how to formulate non-relativistic NCQM in 1+1 dimensional Moyal spacetime, in a user friendly way, which mandates the formulation of an equivalent commutative theory, circumventing the famous Pauli's objection \cite{pau} and also facilitates the study of some physical implications.\\
Moreover, it is quite interesting to point out here that, the authors in \cite{v1} have shown how one can extend the notion of classical space time to that of a quantum space-time, such that the well-known Unruh effect and Hawking radiation, in quantum space-time background can be shown to arise from the relevant nonzero space time commutator. On the other hand, recently in \cite{mann}, the authors have shown a connection between semi-classical quantization of gravity and geometric phase. Furthermore, it has been shown that, due to its motion in space-time, the state of a detector coupled to a quantum field ,  acquires a geometric phase which is a functional of the trajectory of the detector and depends on Unruh temperature. Interestingly, this phase turns out to be within observable limits for values of acceleration as low as $\sim$10 m/$s^2$. So by measuring the geometric phase or namely Berry phase one can experimentally detect Unruh effect even at this low acceleration. At this stage, one may recall that it was M. Berry \cite{berry,berry1} who showed that if the Hamiltonian of a quantum system depends on a number of parameters, varying adiabatically in a cyclic fashion, then the state of the system
acquires an extra phase factor in addition to the usual
dynamical phase. This additional factor, known as Berry phase, depends only on the geometrical structure of the parameter space and not on the time of evolution. This effect was further generalized to arbitrary cyclic evolution
by Aharonov and Anandan \cite{aranov} and this has been tested quite extensively 
for two level systems \cite{tomita}. Returning back to our main point, we note that the result in the previously cited paper \cite{mann} shows that the geometric phase encodes the information of the Unruh temperature indicating the existence of a relation between Berry phase and Unruh effect. Although the universality of Berry phase can be seen in variety of contexts like Born-Oppenheimer approximation \cite{mina}, quantum Hall effect \cite{fuzi,hall,zhang} etc, there is still a gap in literature suggesting any connection between Berry phase and quantum gravity or for that matter in the simplest set-up of quantum space-time of the type we have mentioned above. So it is quite appropriate to study the occurrence of geometric phase, if any, in such a quantum space-time. With this motivation in mind, we undertake the study of the relation between NC space time and emergence of geometric phase in a dynamical system. For that we have constructed the effective commutative Hamiltonian of a forced harmonic oscillator placed in a 2 dimensional quantum (Moyal) space-time, by employing the prescription provided in \cite{partha}. This gives rise to a generalised harmonic oscillator system with perturbations linear in position  and momentum respectively. An adiabatic evolution of the total Hamiltonian is then considered through the slowly varying time dependent coefficients. As in \cite{partha}, here too we have studied the time evolution of the ladder operators (with which the system is diagonalised) in Heisenberg's picture and have shown that the adiabatic evolution of the system over a total time period $\mathcal{T}$ gives rise to a total phase consisting of the usual dynamical phase along with an extra geometric phase shift in the operator, which vanishes in the commutative limit.\\ \\
The paper is organized as follows: In sec-2 we have provided a review of time-reparametrization invariant formulation of non-relativistic particle moving in the commutative space-time and eventual emergence of Schr\"{o}dinger equation as a constraint equation at the quantum level paving the way for its generalization to noncommutative space-time in (1+1) D which we take up in the next sec-3, where we have also shown how the  non-commutativity of space-time appears through the Dirac brackets in a classical toy model and then quantized it. Here we have introduced the Hilbert Schmidt (HS) operatorial formulation to discuss NC quantum mechanics. In sec-4 we have considered a forced harmonic oscillator system in NC space-time and then made use of coherent state basis to obtain an effective commutative Hamiltonian responsible for the time evolution through the effective commutative  Schr\"{o}dinger equation. By carrying out an additional time-dependent unitary transformation the system is transformed to a time-dependent generalised harmonic oscillator (GHO) system. We then consider an adiabatic time evolution, over the time period $\mathcal{T}$, of the ladder operators of the GHO system, which is then shown to yield the geometric i.e. Berry phase along with a dynamical phase in sec-5. Finally we conclude and try to give some future prospects of the work in sec-6. Apart from these sections in main text, we have also included three appendices to complement the materials presented in the text. And they are given as follows: In appendix-A.1, we have shown the effect of space-time noncommutativity on the dynamics of a time \textit{independent} system. Next in appendix-A.2, we have carried out constraint analysis and computed the Dirac brackets for the phase space variables for a particular (1+1) D classical toy model, which produces non-vanishing Dirac brackets between the configuration space variables. Then in appendix-A.3, we provide a detailed derivation of the inner product in the space of square integrable functions $L^2_{\star}(\mathbb{R}^1)$ in the non-commutative space-time, starting from that of usual commutative space-time : $L^2(\mathbb{R}^1)$, where the functions are composed using point-wise multiplication.  Finally in appendix-A.4, we have shown explicit computation of the geometric phase shift for the system of interest. 
\section{A classical picture of non-commutative space-time from time\\ reparametrization symmetry}
In this section we first briefly sketch the idea of time-reparametrization invariant form of classical action.  We introduce a modified Lagrangian formalism (non-relativistic), where the usual commutative time $t$ is treated as an additional configuration space variable and is on an equal footing with the position coordinates \cite{Deri}. \\
We begin by considering the Lagrangian $L^t(x_i,\dot{x}_i,t)$ of a dynamical system in 1+1 dimension and treat time $t$, as the zero-th component of generalised coordinate $x^{\mu}$ : $x^0=t$. Next we introduce a new evolution parameter $\tau$ for the trajectory of the system and for that all we want is to just require that,  $t$ is a monotonically increasing function of $\tau:\,\, t=t(\tau)$. So with the identification $dt = \dot{t} d\tau$ and $\dot{x}_{i}= \left(\frac{dx_i}{d\tau}\right)$, with overdots, from here onwards, representing the derivative with respect to $\tau$, the time-reparametrization invariant form of action can be written as 
\begin{equation}
S= \int \, dt \, L^t\left(x_i,\frac{dx_i}{dt},t\right) =\int\, d\tau \,\,\dot{t} \,L^t (x_i,\dot{x}_i,t,\dot{t}) =\int \,\,d\tau\,\, L^{\tau}(x^{\mu},\dot{x}^{\mu})\label{e53}
\end{equation}  
which remains invariant under  the world line reparametrization $\tau \to \tau ' =\tau '(\tau),\,\,\,\,$ with $x^{\mu}(\tau)$ transforming as world-line scalars: $ x^{\mu} (\tau)\,\to\, x^{' \mu}(\tau ') = x^{\mu} (\tau)$ under this reparametrization. Note that we have identified $L^{\tau}(x^{\mu},\dot{x}^{\mu})$ the integrand in (\ref{e53}) as our modified Lagrangian. At this point let us consider the simple example of  Lagrangian of a non-relativistic particle in presence of a potential $V(x,t)$:
\begin{equation}
L^t(x,\dot{x},t)= \frac{1}{2}m\left(\frac{dx}{dt}\right)^2-V(x,t)\label{e55}
\end{equation}
 Considering the evolution parameter to be $\tau$ , the corresponding modified Lagrangian of the theory is obtained as,
\begin{equation}
L^{\tau}= \frac{1}{2} m\frac{\dot{x}^2}{\dot{t}}-\dot{t}V(x,t)\label{e56}
\end{equation}
The canonical momenta corresponding to the generalised coordinates $x(\tau)$ and $t(\tau)$ are now given by,
\begin{align}
p_x&=\frac{dL^{\tau}}{d\dot{x}}=m\frac{\dot{x}}{\dot{t}}=m\frac{dx}{dt}\nonumber\\
p_t&=\frac{dL^{\tau}}{d\dot{t}}=-\frac{m}{2}\left(\frac{dx}{dt}\right)^2 -V(x,t)=-\frac{p_x^2}{2m}-V(x,t)=-H\label{e54}
\end{align}
where $H= \frac{p_x^2}{2m}+V(x,t)$ is the canonical Hamiltonian corresponding to (\ref{e55}). The above equation (\ref{e54}) indicates the presence of a primary constraint as
\begin{equation}
\phi =p_t+H \approx 0\label{e57}
\end{equation}
where the notation $\approx$ is used in sense of weak equality a \textit{la} Dirac \cite{dir}. The new Hamiltonian corresponding to the redefined Lagrangian $L^{\tau}$ (\ref{e56}) is proportional to $\phi$ (\ref{e57}) and therefore also vanishes weakly:
\begin{equation}
	H^{\tau}= p_t \dot{t}+p_x\dot{x}-L^{\tau}=\dot{t}\phi\approx 0\label{e76}
\end{equation}
So we see that the system is endowed with a first class constraint $\phi$. The total Hamiltonian according to Dirac's prescription for constrained theory is then given by
\begin{equation}
H_T^{\tau}= 	H^{\tau}+\sigma(\tau)\phi =\sigma(\tau)\phi
\end{equation}
$\sigma(\tau)$ being the Lagrange's multiplier. We can now write the Lagrangian in the first order form given by the inverse Legendre transformation as
\begin{equation}
L_f^{\tau}=p_{\mu}\dot{x}^{\mu}-\sigma(\tau)(p_t+H),\,\,\,\,\,\mu=0,1\label{e65}
\end{equation}
Note that in the  first order formalism, the configuration space coincides with the phase space i.e. the configuration space is enlarged further by taking the canonical momenta also as generalised coordinates. Now we can easily run through the algorithm of Dirac's constraint analysis \cite{dir} for the above Lagrangian (\ref{e65}) to find one primary first class constraint and two pairs of primary second class constraints. The second class constraints of the system are then implemented strongly by introducing  Dirac's brackets (DB), which can be computed easily and finally the DBs among the phase space variables are now obtained as:
\begin{equation}
\{x^{\mu},x^{\nu}\}_D=0=\{p_{\mu},p_{\nu}\}_D;\,\,\,\,\,\,\{x^{\mu},p_{\nu}\}_D=\delta^{\mu}\,_{\nu}
\end{equation}
Canonical quantization is now carried out by elevating the phase space variables ($t,x,p_t,p_x$) to the level of operators and elevating the Dirac brackets to the level of commutator brackets as
\begin{equation}
[\hat{t},\hat{x}]=0=[\hat{p}_t,\hat{p}_x],\,\,[\hat{t},\hat{p}_t]=i=[\hat{x},\hat{p}_x] \,\,\,\,(\hbar=1)\label{e60}
\end{equation}
We then look for an appropriate Hilbert space on which these operators act and furnish a suitable representation of this algebra. The quantum counterpart of the 2D configuration space is now the Hilbert space $L^2(\mathbb{R}^2)$. We now introduce the simultaneous ``spatio-temporal" eigenbasis $\left|x,t\right\rangle$ of the commutating  operators $\hat{t}$ and $\hat{x}$ satisfying
\begin{equation}
\hat{t}\left|x,t\right\rangle=t\left|x,t\right\rangle,~~~\hat{x}\left|x,t\right\rangle=x\left|x,t\right\rangle.\label{A3}
\end{equation}
which fulfills the following completeness and orthonormality relations :
\begin{equation}
\int dt dx \left|x,t \right\rangle \langle x,t | =\mathbb{I}, ~~~\left\langle x,t | t^{'},x^{'} \right\rangle=\delta(t-t^{'})\delta(x-x^{'}).
\end{equation}
The coordinate representations of phase space operators are given by
\begin{equation} \label{n2}
\begin{split}
\left\langle x,t | \hat{x} | \psi \right\rangle = x\left\langle x,t |\psi \right\rangle , & \left\langle x,t |\hat{t} |\psi \right\rangle = t \left\langle x,t |\psi \right\rangle\\
\left\langle x,t |\hat{p}_x |\psi \right\rangle = -i\partial_x \left\langle x,t |\psi \right\rangle , & \left\langle x,t |\hat{p}_t |\psi \right\rangle = -i\partial_t \left\langle x,t |\psi \right\rangle \
\end{split}
\end{equation}
where $\psi (x,t)=\left\langle x,t|\psi\right\rangle \in L^2(\mathbb{R}^2)$ and can be formally identified with the function satisfying the following square-integrable norm:
\begin{equation}
\left\langle\psi |\psi\right\rangle =\int dtdx~ \psi^\ast(x,t)\psi(x,t) < \infty .\label{nn2}
\end{equation}
The associated inner product is given by
\begin{equation}
<\psi|\phi>_=\int dt dx \psi^{\star}(x,t)\phi(x,t)
\label{in2}
\end{equation}
In order to obtain an appropriate probabilistic interpretation of non-relativistic QM, however, we need to consider a physical Hilbert space which is the linear span of all possible physical states and can be identified by projecting out physical states ($\left|\psi_{phy}\right\rangle$) by imposing the subsidiary condition
\begin{equation}
\hat{\phi}\left|\psi_{phy}\right\rangle=(\hat{p_t} + \hat{H})\left|\psi_{phy}\right\rangle=0,\label{nn1}
\end{equation}
where $\hat{\phi}$ is the operator form of the first class constraint (\ref{e57}) \cite{Deri}. This is tantamount to demanding the gauge invariance  of physical states.\\
The coordinate representation of quantum constraint equation (\ref{nn1})  readily yields the time evolution of the physical states or wave-functions ($<x,t|\psi_{phy} >:=\psi_{phy}(x,t)$) as
\begin{equation}
i\frac{\partial}{\partial t} \psi_{phy}(x,t)=\left( -\frac{1}{2m} \frac{\partial^2}{\partial x^2} + V(x,t)\right) \psi_{phy}(x,t)\label{A7}
\end{equation}
which we recognize as the time-dependent Schr\"odinger equation. Note that it is independent of the parameter $\tau$, as its $\tau$-evolution, which is nothing but the unfolding of gauge transformation generated by the first class constraint $\phi$ (\ref{e57}), is now frozen, as can be easily observed by using (\ref{e76},\ref{nn1}).\\
Now to extract the probabilistic interpretation, we recall that the continuity equation corresponding to this Schr\"odinger equation, given by
\begin{equation}
\frac{\partial \rho}{\partial t}+\frac{\partial J_{x}}{\partial x}=0
\label{conti}
\end{equation}
with $\rho=\psi^{\star}_{phy}(x,t)\psi_{phy}(x,t)$ and $J_{x}=\frac{1}{2m}Im(\psi^{\star}_{phy}(x,t)\overleftrightarrow{\partial_{x}}\psi_{phy}(x,t))$. Correspondingly, after performing a spatial integration in both sides of (\ref{conti}) ranging from -$\infty$ to +$\infty$ we can write,
\begin{equation}
\partial_t\int_{-\infty}^{\infty} \rho dx =-\int_{-\infty}^{\infty} (\partial_xJ_x) dx= 0
\end{equation}
where we have used the fact that our physical wave-function satisfies $\psi_{phy}(x,t)\to 0$ as $x\to \pm\infty$ unlike in (\ref{nn2}) where we also necessarily require $\psi_{phy}(x,t) \to 0$ as $t \to \pm\infty$, which clearly will not be suitable for a probabilistic interpretation. Therefore the nonnegative $\rho$ can indeed be interpreted as a probability density and this notion of physical state defined on a ``space-like surface" indicates the conservation of the total probability $\int_{-\infty}^{\infty} \rho dx$. From the expression of the continuity equation (\ref{conti}), the variable $t$ can be identified with time which now plays the role of the evolution parameter. We thus see that  for $\psi_{phy}(x,t)$ to be the so called  \textquotedblleft{well} behaved" it has to be normalizable and must satisfy the following square integrability condition at a constant time slice as 
\begin{equation}
\langle \psi_{phy} |\psi_{phy} \rangle_{t} =	\int_{-\infty}^{\infty} \,dx\,\, \psi_{phy}^*(x,t)\psi_{phy}(x,t) < \infty\label{e44}
\end{equation}
In other words, the physical states satisfying (\ref{A7}) cannot be elements of the Hilbert space $L^2(\mathbb{R}^2)$. We shall, however, restrict our attention to physical states and we shall henceforth suppress \textquoteleft{phy}" in the subscript and simply write it as $\psi(x,t)$ for brevity and will insert it back in the subscript as and when required explicitly. More precisely, the probabilistic notion in QM is recovered by replacing the inner-product (\ref{in2}) by the one, which involves only a spatial integration at a constant time slice:
\begin{equation}
\left\langle\psi |\phi\right\rangle_t : =\int_{t} dx ~\psi^\ast(x,t)\phi(x,t).\label{in1}
\end{equation}
We shall refer this as ``induced inner product". Clearly, any normalizable states with $L^2(\mathbb{R}^1)$ inner product (\ref{in1}) may not be so with respect to that of $L^2(\mathbb{R}^2)$ (\ref{in2}).\\ \\
As an example, one may consider a stationary state, associated with energy $E$, of the form $\psi_E(x,t)=e^{-iEt}\psi(x)$. Clearly, this $\psi_E(x,t) \in L^2(\mathbb{R}^1)$ but $\psi_E(x,t) \notin L^2(\mathbb{R}^2)$. Further, note that this state can be written as a superposition of plane waves corresponding to different momentum $p$ as $\psi_E(x,t)= \int\,dp\, e^{-i(Et-px)}\tilde{\psi}(p)$, where the exponential $\frac{1}{2\pi}e^{-i(Et-px)}:= \langle x,t|E,p\rangle $ is the overlap of simultaneous eigenstate $|E,p\rangle $ of both $\hat{p}_t$ and $\hat{p}_x$ : $\hat{p}_t|E,p\rangle =- E |E,p\rangle $ and $\hat{p}_x |E,p\rangle = p|E,p\rangle $, with the basis $|x,t\rangle $ (\ref{A3}). Now if we consider a pair of such stationary states $|E,p\rangle $ and $|E',p'\rangle $, then using the original inner product (\ref{in2}), one gets  
\begin{equation}
	\langle E',p'|E,p\rangle =\delta(E'-E)\,\delta (p'-p)\label{M2}
\end{equation}
This clearly shows that for coincident energy eigenvalues i.e. $E'=E$, the right hand side in (\ref{M2}) diverges. In contrast, the induced inner product (\ref{in1}) in this case yields
\begin{equation}
	\langle E,p'|E,p\rangle _t= \frac{1}{2\pi}\delta(p'-p)\label{M3}
\end{equation}
which is nothing but the usual orthonormality condition for momentum eigenstates which is free of such divergence. As we shall demonstrate in the sequel that an exactly similar situation arises in our kind of NC space-time as well.
In either case, therefore, we shall primarily be working with this induced inner product or more generally between pair of wave packets like $|\psi\rangle $ and $|\phi\rangle $ (\ref{in1}), so that we can recover the probabilistic interpretation using this inner product of $L^2(\mathbb{R}^1)$ only.\\ \\
  Finally, note that the self-adjoint property of the derivative representation of $\hat{p}_t=-i\partial_t$ in (\ref{n2}) does not hold anymore in the Hilbert space $L^2(\mathbb{R}^1)$ with associated inner product (\ref{in1}), as it is impossible to demand that $\mid\psi(x,t)\mid \to 0$ as $\mid t\mid \to \infty$ as mentioned above, which is required in $L^2(\mathbb{R}^2)$, to allow one to carry out integration by parts with respect to $'t'$ and drop boundary contribution. We shall thus exclude $\hat{p}_t$  from the dynamical phase-space variables, along with $\hat{t}$. The latter, when `demoted' to a $c$-number parameter, is now identified with the new evolution parameter with $\left( -i\partial_t \right)$ having no association with $\hat{p}_t$ anymore, so that (\ref{A7}) has now the status of a postulate.
  \section{Quantum mechanics on quantum space-time}
It has already been shown in \cite{basu} that the study of a typical time independent system on NC space produces new dynamics of the system. The form of the propagator of the system, gets modified and we also get a correction in the wave function of the system. On the other hand, when such a system is placed in a NC space-time (i.e. with time also being an operator) background, the system dynamics gives equivalent result (see Appendix-A.1) and no NC correction is obtained in the Hamiltonian and therefore in the spectrum of the system. This motivates us to examine a time \textit{dependent} system placed in a NC space time and look for signatures, if any, of this noncommutativity on the  dynamics of the system. And in this regard, the discussion on QM on commutative space-time in the preceding section, will play an important role as we shall be basically emulating the same template to formulate the QM in our (1+1) D Moyal plane. And for that we consider a time dependent Hamiltonian of a forced harmonic oscillator (FHO) which is known to have many applications in quantum optics. This was primarily motivated by the structure of the Hamiltonian, proposed by Mehta and Sudarshan \cite{mehta}, where it is shown that coherent states persist to be coherent after time evolution, if we require the general form of the Hamiltonian to be that of a FHO. Also in \cite{caru} authors have taken similar Hamiltonian to show that a FHO coupled to a transient classical force can easily be described in terms of coherent states, which is very useful in describing light from optical masers. Thus our primary goal is to look for the above mentioned signature, in the form of geometric phase in such system and for that we first need to  set up the formalism of QM itself of NC space time with time also being an operator. We intend to deal with this in this sub-section. For that  we will first very briefly demonstrate how the NC configuration space can appear in the context of a classical dynamical system.\\ \\
 As indicated previously we have an enlarged configuration space in the first order formulation, which allows more freedom for theoretical interest. In our purpose, we can add a Chern-Simon (CS) like term in the momentum space to the existing first order Lagrangian $L^{\tau}_f$ (\ref{e65}) \cite{luk}. This is motivated from \cite{sg,pn}, where it was shown that the phase-space representation of action, of a particle placed in a NC plane in presence of an arbitrary potential gives rise to an additional \textquotedblleft{exotic}" term in action, which is a Chern-Simon (CS) like term in momentum space and the corresponding CS parameter is essentially of NC origin. Also the presence of this CS term in the Lagrangian does not affect the Gallelian symmetry of the system.\footnote{For a Gallelian transformation given by $\delta x= vt, \delta t=0$ we get the transformation in the following momentum as : $\delta p_x= mv, \delta p_t =-\delta H_{free}=-p_x v $ where $H_{free}$ is the free particle Hamiltonian. With this transformation we can easily verify that the Chern-Simon term $\epsilon_{\mu\nu}p_{\mu}\dot{p}_{\nu}$  transforms up to a total derivative which can be neglected in the action. So the CS action is quasi invariant under Gallelian transformation. }\\
So we now take the following Lagrangian to obtain a NC generalisation of the commutative space-time in a classical setting,
\begin{equation}
L_f^{\tau,\theta}= p_{\mu}\dot{x}^{\mu}+\frac{\theta}{2}\epsilon^{\mu\nu}p_{\mu}\dot{p}_{\nu}-\sigma(\tau)(p_t+H),\,\,\,\,\,\mu ,\nu = 0,1\label{e73}
\end{equation}
where $p_{\mu}=(p_t,p_x)$ .
We can again run the Dirac's algorithm (Appendix-A.2) to get the following Dirac brackets between the phase space variables.
\begin{equation}
\{x^{\mu},x^{\nu}\}_D=\theta\epsilon^{\mu\nu};\,\,\,\,\,\{p_{\mu},p_{\nu}\}_D=0;\,\,\,\,\,\,\{x^{\mu},p_{\nu}\}_D=\delta^{\mu}\,_{\nu}\label{lev}
\end{equation} 
So we can see that the NC nature of configuration space emerges very naturally from a constrained classical system. Note that this is shown just to motivate the appearance of non-commutativity between coordinate variables naturally at the classical level before setting up a quantum mechanical description of the quantum space-time. In this paper we shall, of course, be dealing with a different dynamical system in the next section.\\
Now to provide a description for the non-relativistic QM in quantum space-time, we will elevate the classical brackets (\ref{lev}) to the level of commutation brackets as given below, 
\begin{equation}
	[\hat{t},\hat{x}]=i\theta\label{e1}
	\end{equation} 
with $\theta$ being the NC parameter, along with 
\begin{equation}
[\hat{p}_t,\hat{p}_x]=0,\,\,[\hat{t},\hat{p}_t]=i=[\hat{x},\hat{p}_x].\label{e2}
\end{equation} 
Recall that we are working in the natural unit $\hbar= 1$ throughout this paper. (\ref{e1}) and (\ref{e2}) as a whole represents the NC Heisenberg algebra (NCHA). \\ 
 A representation of NC  coordinate sub-algebra in (\ref{e1}) is furnished by the Hilbert space  \cite{partha,biswa,pnskp,gauba},
\begin{equation}
\mathcal{H}_c=Span \left\{|n\rangle = \frac{(b^{\dagger})^n}{\sqrt{n!}}|0\rangle;\,\,b|0\rangle =\frac{\hat{t}+i\hat{x}}{\sqrt{2\theta}}|0\rangle =0 \right\}\label{e3}
\end{equation} 
Let us now introduce the associative NC operator algebra ($\hat{\mathcal{A}}_{\theta}$) generated by ($\hat{t},\hat{x}$) or equivalently  ($\hat{b},\hat{b}^{\dagger}$)  acting on this configuration space $\mathcal{H}_c$ (\ref{e3}) as
\begin{equation}
\hat{\mathcal{A}}_{\theta}=\{|\psi)=\psi (\hat{t},\hat{x})=\psi(\hat{b},\hat{b}^{\dagger})= \sum_{m,n} c_{n,m}|m\rangle\langle n|\}\label{e59}
\end{equation}
which is basically the set of all equivalence class of polynomials in $(\hat{t},\hat{x})$ or ($\hat{b},\hat{b}^{\dagger}$), subject to the identification $[\hat{b},\hat{b}^{\dagger}]=1$, which also can be thought of as the universal enveloping algebra corresponding to (\ref{e1}). At this stage one should note that $\hat{\mathcal{A}}_{\theta}$ has not yet been endowed with any inner product and norm structure and consequently can't be identified with any Hilbert or Banach space at this stage.\\ 
We can now introduce a subspace $\mathcal{H}_q \subset \mathcal{B}(\mathcal{H}_c) \subset \hat{\mathcal{A}}_{\theta}$ of $\hat{\mathcal{A}}_{\theta}$ (\ref{e59}) as the space of `Hilbert Schmidt' (HS) operators, which are bounded and compact operators with finite HS norm, again acting on $\mathcal{H}_c$ (\ref{e3}), and is given by,
\begin{equation}
\mathcal{H}_q= \left\{\psi(\hat{t},\hat{x})\equiv \Big|\psi(\hat{t},\hat{x})\Big)\in \mathcal{B}(\mathcal{H}_c) ;\, \, ||\psi||_{HS}:=\sqrt{tr_c(\psi^{\dagger}\psi)} < \infty\right\} \subset \hat{\mathcal{A}}_{\theta}\label{e4}
\end{equation}
where $||\psi||_{HS}$ is the Hilbert-Schmidt norm and $tr_c$ denotes trace over the Hilbert space $\mathcal{
H}_c$ and $\mathcal{B}(\mathcal{H}_{c})\subset \hat{\mathcal{A}}_{\theta}$ is a set of bounded operators on $\mathcal{H}_{c}$.  This space is equipped with the inner product
\begin{equation}
\Big(\psi(\hat{t},\hat{x}),\phi(\hat{t},\hat{x})\Big):=tr_{c}\Big(\psi^{\dagger}(\hat{t},\hat{x})\phi(\hat{t},\hat{x})\Big)
\label{iop}
\end{equation}
and so forms a Hilbert space on its own.  Note that as a matter of notations the elements of $\mathcal{H}_c$ and $\hat{\mathcal{A}}_{\theta}$ are denoted by the angular ket $| .\rangle$ and round ket $| . )$ respectively. We now define the quantum space-time coordinates ($\hat{T},\hat{X}$) and corresponding conjugate momentum operators ($\hat{P_{t}},\hat{P}_{x}$)  by their actions on a typical element $\Big|\psi(\hat{t},\hat{x})\Big)\in \mathcal{H}_{q}$ as,
\begin{align}
&\hat{T}\Big|\psi(\hat{t},\hat{x})\Big)=\Big|\hat{t}\psi(\hat{t},\hat{x})\Big),\,\,\,\,\,\hat{X}\Big|\psi(\hat{t},\hat{x})\Big)=\Big|\hat{x}\psi(\hat{t},\hat{x})\Big),\nonumber\\
&\hat{P}_x\Big|\psi(\hat{t},\hat{x})\Big)=-\frac{1}{\theta}\Big|[\hat{t},\psi(\hat{t},\hat{x})]\Big),\,\,\,\hat{P}_t\Big|\psi(\hat{t},\hat{x})\Big)=\frac{1}{\theta}\Big|[\hat{x},\psi(\hat{t},\hat{x})]\Big) \label{e5}
\end{align}
Note that here upper case letters $\hat{T}$ and $\hat{X}$ are used to distinguish them from their the lower case counterparts $\hat{t}$ and $\hat{x}$, as their domain of actions are different; while $(\hat{t},\hat{x})$ act on $\mathcal{H}_{c}$,  $(\hat{T},\hat{X})$ act on $\mathcal{H}_{q}$. But since they satisfy isomorphic commutator algebra:$[\hat{T},\hat{X}]=i\theta$, $(\hat{T},\hat{X})$ can be regarded as the representation of ($\hat{t}, \hat{x}$). It is easily verified using (\ref{e5}) that ($ \hat{T},\hat{X}, \hat{P}_{t},\hat{P}_{x}$) satisfy the same NC Heisenberg algebra (\ref{e1},\ref{e2}). These new-fashioned operators can be thought as \textquotedblleft{super} operators", as they operate again on space of operators. In fact, they can act on the entire $\hat{\mathcal{A}}_{\theta}$. \\
However, it is clear from the discussion of the previous section that HS norm (\ref{e4}) has its commutative counterparts in (\ref{nn2}). Clearly, the associated inner-product (\ref{iop}), like the counterpart (\ref{in2}), needs also to be modified to some appropriate form analogous to (\ref{in1}) if we want to have a proper probabilistic interpretation. We take it up in the next sub-section.
\subsubsection{Schr\"odinger equation and an induced inner product}
Now it is clear that in view of $\theta \neq 0$, that we cannot find a counterpart of the common space-time eigenstate $|x,t\rangle $ (\ref{A3}). However, we can still recover an effective commutative theory by making use of the coherent state. For that we choose the Sudarshan-Glauber coherent state made out of the Fock states $|n\rangle$ belonging to $\mathcal{H}_c$ (\ref{e3}) as
 \begin{equation}
 	|z\rangle = e^{-\bar{z}\tilde{b}+z\tilde{b}^{\dagger}}|0\rangle\,\in\mathcal{H}_c\label{e43}
 \end{equation} 
 which is an eigen state of the annihilation operator : $\tilde{b}|z\rangle=z|z\rangle$, where $\tilde{b}=\frac{\hat{t}+i\hat{x}}{\sqrt{2\theta}}$ and $z$ is dimensionless complex number and is given by,
 \begin{equation}
 	z=\frac{t+ix}{\sqrt{2\theta}};\,\,\,\,\,\,\,\,\,t=\langle z|\hat{t}|z\rangle , x=\langle z|\hat{x}|z\rangle\label{A4}
 \end{equation}
  Here $t$ and $x$ are effective commutative coordinate variables. We can now construct the counterpart of coherent state basis in $\mathcal{H}_q$ (\ref{e4}), made out of the bases $|z\rangle \equiv |x,t\rangle$ (\ref{e43}), by taking their outer product as
 \begin{equation}
 |z,\bar{z})\equiv |z)=|z\rangle\langle z|=\sqrt{2\pi \theta}\,\,|x,t)\,\in \mathcal{H}_q \,\,\,\,\,\,\textrm{fulfilling}\,\,\,B|z)=z|z)\label{A8}
 \end{equation}
 where the annihilation operator $\hat{B}=\frac{\hat{T}+i\hat{X}}{\sqrt{2\theta}}$ is a representation of the operator $\tilde{b}$ in $\mathcal{H}_q$ (\ref{e4}). In this context, note that $|z) \in \mathcal{H}_q$ (\ref{A8}) saturates the space-time uncertainty: $\Delta \hat{T}\, \Delta \hat{X}= \frac{\theta}{2}$, implying that such a state represents a maximally localized \textquotedblleft{point}", or rather an event in space-time. In fact this state being a pure density matrix can be regarded as a pure state of the algebra $\hat{\mathcal{A}}_{\theta}$ (\ref{e59}) and plays the role of a point, represented by Dirac's delta functional, in the corresponding commutative algebra $C^{\infty}(\mathbb{R}^2)$ describing (1+1) D commutative plane \cite{chaoba,anwe}. \\
 It can also be checked that the basis $|z,\bar{z})\equiv |z)$ satisfies the over-completeness property:
 \begin{equation}
 \int \,\,\frac{d^2z}{\pi}|z,\bar{z})\,\star_{V}\,(z,\bar{z}|= \int dtdx \, |x,t) \star_{V} (x,t| = \textbf{1}_q,
 \label{vnc}
 \end{equation} \\
 where $*_V$ represents the Voros star product and is given by,
 \begin{equation}
 \star_V=e^{\overleftarrow{\partial_z}\overrightarrow{\partial_{\bar{z}}}}=e^{\frac{i\theta}{2}(-i\delta_{ij}+\epsilon_{ij})\overleftarrow{\partial_i}\overrightarrow{\partial_j}};\,\,\,\,i,j=0,1;\,\,\,x^0=t, x^1=x;\,\,\,\,\,\epsilon_{01}=1
 \end{equation}
 Then the coherent state representation of an abstract state $\psi(\hat{t},\hat{x})$ gives the usual coordinate representation of a state just like formal QM:
 \begin{equation}
 \psi(x,t)=\frac{1}{\sqrt{2\pi\theta}}\Big(z,\bar{z}|\psi(\hat{x},\hat{t})\Big)=\frac{1}{\sqrt{2\pi\theta}}tr_{c}\Big[|z\rangle\langle z|\psi(\hat{x},\hat{t})\Big]=\frac{1}{\sqrt{2\pi\theta}}\langle z|\psi(\hat{x},\hat{t})|z\rangle\label{e45}
 \end{equation}
 and is called the symbol of the HS operator $\psi(\hat{x},\hat{t})$.
 The corresponding representation of a composite operator say $\psi(\hat{x},\hat{t}) \phi(\hat{x},\hat{t})$ is  given by composing the corresponding symbols through Voros star product and is given as 
 \begin{equation}
 \Big(z\Big|\psi(\hat{x},\hat{t})\phi(\hat{x},\hat{t})\Big)=\Big(z\Big|\psi(\hat{x},\hat{t})\Big) \,\star_V\,\Big(z\Big|\phi(\hat{x},\hat{t})\Big)
 \label{como}
 \end{equation} 
 This establishes an isomorphism between the space of HS operators $\mathcal{H}_q$ and the space of their respective symbols, whereas in the former case, the composition rule is given by the usual product of operators, in the latter case it is given by the Voros star product.  Now,  using (\ref{vnc})  the overlap of two arbitrary states ($|\psi),|\phi)$) in the quantum Hilbert space $\mathcal{H}_q$ can be written in the form
 \begin{equation} 
 (\psi|\phi) = \int dtdx ~ \psi^\ast(x,t) \star_{V} \phi(x,t)\label{innpro_voros}
 \end{equation}
 We shall demonstrate in the sequel that with this Voros star product probability density is indeed positive definite. 
 Also the basis $|z,\bar{z})_V$ which we can refer to as Voros basis as in \cite{basu} is compatible with POVM (Positive operator valued measure) \cite{povm} in contrast to Moyal basis which is associated to a similar kind of star product named as Moyal star product \cite{basu} in the context of (2+1) D QM in the Moyal plane, where the time was not an operator. Henceforth, we shall thus be using only this Voros basis and the associated Voros star product for our purpose, so we shall omit the subscript V from now onwards. \\
 Now to obtain the effective commutative Schr\"odinger equation in coordinate space, we introduce coordinate  representation of the phase space operators. To begin with note that the coherent state representation of the left actions of space-time operators  $\{\hat{X}_{L},\hat{T}_{L}\}$, acting on any arbitrary element $|\psi) \in \mathcal{H}_q$, can be written by using (\ref{como}) as,
  \begin{equation}
 \Big( x,t\Big|\hat{X}_L \, \psi(\hat{x},\hat{t})\Big) =\frac{1}{\sqrt{2\pi\theta}}\Big(z,\bar{z}\Big|\hat{x}\psi\Big) = \frac{1}{\sqrt{2\pi\theta}} \, \left\langle z|\hat{x}|z\right\rangle \star_{V} (z,\bar{z}|\psi(\hat{x},\hat{t}))
 \end{equation}
 Finally making use the fact of \eqref{e45} this readily yields
 \begin{equation}
 \Big( x,t|\hat{X}_L \, \psi(\hat{x},\hat{t})\Big) = X_\theta^L \, \Big(x,t|\psi(\hat{x},\hat{t})\Big) = X_\theta^L \, \psi(x,t)
 \end{equation}
 with
 \begin{equation}
  X_\theta^L =  \left[x+\frac{\theta}{2}(\partial_x-i\partial_t)\right]\label{47}
 \end{equation}
  Proceeding exactly in the same way, we obtain the representation of $\hat{T}$ as
 \begin{equation}
 T_\theta^L =  \left[t+\frac{\theta}{2}(\partial_t+i\partial_x)\right],\label{48}
 \end{equation}
 so that $[T_{\theta}^L,X_{\theta}^L]=i\theta$ is trivially  satisfied when their actions on arbitrary test functions are considered. Now, it is trivial to prove the self-adjointness property of both $X_\theta$ and $T_\theta$, w.r.t. the inner product \eqref{innpro_voros} in $\mathcal{H}_{q}$ by considering an arbitrary pair of different states $|\psi_1)$, $|\psi_2) \in \mathcal{H}_q$ and their associated symbols, just by exploiting associativity of Voros star product. We also note that since this analysis will not involve any integration by parts i.e. it is not sensitive to the integration measure, this self-adjointness property of $X_\theta$ and $T_\theta$ will persist to hold for the `induced' inner product, to be introduced in the sequel in \eqref{op1} (see below) - the counterpart of (\ref{in1}),  as well. Like-wise the right action of $\hat{X}_R$ and $\hat{T}_R$ on $|\psi)$ defined as
 \begin{equation} 
 \hat{X}_R|\psi) =\Big|\psi(\hat{x},\hat{t})\hat{x}\Big);\,\,\,\,\,\,\hat{T}_R|\psi )= \Big|\psi(\hat{x},\hat{t})\hat{t}\Big)\label{A1}
 \end{equation}
where the corresponding representations are obtained as,
 \begin{equation}
 	X_{\theta}^R=\left[x+\frac{\theta}{2}(\partial_x+i\partial_t)\right],\,\,\,\,\,\,\,\,T_{\theta}^R=\left[t+\frac{\theta}{2}(\partial_t-i\partial_x)\right]\label{A2}
 \end{equation}\\
 fulfilling $[T_{\theta}^R,X_{\theta}^R]=-i\theta$, appropriate for the generators of opposite algebra $\hat{\mathcal{A}}_{\theta}^o$.\\
 Finally coming to the coherent state representation of momenta operators in (\ref{e5}) we observe that since momenta operators act adjointly, their actions can be written equivalently as the corresponding differences between left and right actions yielding, 
  \begin{equation}
 \Big(x,t|\hat{P}_t \psi(\hat{x},\hat{t})\Big)= -i \partial_t\psi(x,t)~;~~ \Big(x,t|\hat{P}_x \psi(\hat{x},\hat{t})\Big)=-i\partial_x\psi(x,t)
 \label{25}
 \end{equation}
 Now to isolate the physical Hilbert space $\mathcal{H}_{ph} \subset \hat{\mathcal{A}}_{\theta}$, we note that  our NC theory (\ref{e73}) is endowed with a secondary first-class constraint  $\phi\approx 0$ (\ref{e87}) (see Appendix-A.2). Thus, to obtain an effective commutative Schr\"odinger equation in NC space-time we must impose the condition that the physical states $|\psi_{phy})=\psi_{phy}(\hat{x},\hat{t})$ are annihilated by the operatorial version of this constraint $\phi$:
  \begin{equation}
 (\hat{P}_t+\hat{H})|\psi_{phy})=0;\qquad\psi_{phy}(\hat{x},\hat{t})\in \hat{\mathcal{A}}_{\theta} \label{e74}
 \end{equation} 
 where $\hat{H}=\frac{\hat{P}_x^2}{2m}+V(\hat{X},\hat{T})$. This is just the counterpart of (\ref{nn1}) in the commutative case.
 We are now ready to write down the effective commutative time dependent Schr\"{o}dinger equation in quantum space-time by taking the representation of (\ref{e74}) in $|x,t)$ basis.
 Using (\ref{47},\ref{48},\ref{25}) we finally get,
 \begin{equation}
 i\partial_t \psi_{phy}(x,t)= \left[-\frac{1}{2m}\partial_x^2+ V(x,t)\, \star_{V}\right] \psi_{phy}(x,t)\label{e70}
 \end{equation}
 Taking complex conjugate of this Schrodinger equation we get
 \begin{equation}
 -i\partial_t \psi_{phy}^*(x,t)= -\frac{1}{2m}\partial_x^2 \psi_{phy}^*(x,t) + \psi_{phy}^{*}(x,t)\ \star_{V} V(x,t).\label{e71}
 \end{equation}
 Now using (\ref{e70}) and (\ref{e71}) one therefore obtain the continuity equation
 \begin{equation}
 \partial_t \rho_{\theta} +\partial_x J_{\theta}^{x}=0\label{e72}
 \end{equation}
 where 
 \begin{align}
 \rho_{\theta} &= \psi_{phy}^*(x,t)\,\star_{V}\,\psi_{phy}(x,t),\nonumber\\
 J_{\theta}^{x}&= \frac{1}{2im}[\psi_{phy}^*\star_{V}(\partial_x \psi_{phy})-(\partial_x\psi_{phy}^*)\star_{V}\psi_{phy}].
 \label{prob}
 \end{align}
A spatial integration over $x$ on both sides of (\ref{e72}) over the entire range of $\mathbb{R}^1$ now yields,
 \begin{equation}
 \partial_t\int_{-\infty}^{\infty} \rho_{\theta} dx =-\int_{-\infty}^{\infty} (\partial_xJ_\theta^{\theta}) dx=0
 \end{equation}
 where we have imposed the following condition on $\psi_{phy}$: $\psi_{phy}(x,t) ,\to 0$ as $x\to \pm\infty$ i.e. $\psi_{phy}(x,t) \in L^1_*(\mathbb{R}^1)$ is square integrable \footnote{It may be noted that    $L_{\star}^2(\mathbb{R}^1)$ involves $\theta$-deformed $\star$ multiplication rule, in contest to $L^2(\mathbb{R}^1)$ which involves just the usual point-wise multiplication.}, so that right hand side of the above equation becomes zero giving the conservation of the total probability $\int_{-\infty}^{\infty} \rho_{\theta} dx$ whereby we can write $\rho_{\theta}(x,t)$ in a manifestly positive definite form.
 \begin{equation} 
 \rho_{\theta}(x,t) = \psi_{phy}^*(x,t)\,\star_{V}\,\psi_{phy}(x,t)= \frac{1}{2\pi\theta} \psi_{phy}^\ast(z,\bar{z})\star_V \psi_{phy}(z,\bar{z}) = \frac{1}{2\pi\theta} \sum_{n=0}^\infty \frac{1}{n!}|\partial_z^n \psi_{phy}(z,\bar{z})|^2 \geqslant 0.\label{e75}
 \end{equation} 
 In view of the non-negative condition \eqref{e75} $\rho_{\theta}(x,t)$ can indeed be interpreted as probability density at a particular time. It is clear from the continuity equation (\ref{e72}), the variable $t$ here plays the role of the evolution parameter and for that we require that $\psi_{phy}(x,t)$ to be \textquotedblleft{well} behaved" i.e. it fulfills the following square integrability condition at a constant time slice:
 \begin{equation}
 \langle \psi_{phy} |\psi_{phy} \rangle_{*\,t} =	\int_{-\infty}^{\infty} \,dx\,\, \psi_{phy}^*(x,t)\star_{V}\psi_{phy}(x,t) < \infty,
 \label{op}
 \end{equation}
 so that $\psi_{phy}(x,t)\in L_{\star}^2(\mathbb{R}^1)$  which is quite distinct from $L_{\star}^2(\mathbb{R}^2)$. All of these are straightforward generalization of the corresponding commutative case, discussed in sec-2. See, particularly, below (\ref{in1}). Clearly, the corresponding inner-product for the space should also be defined by a spatial i.e. $x$-integration for a fixed time slice as,
 \begin{equation}
 \langle \psi_{phy} |\phi_{phy} \rangle_{*,\,t} =	\int_{-\infty}^{\infty} \,dx\,\, \psi_{phy}^*(x,t)\star_{V}\phi_{phy}(x,t) < \infty.
 \label{op1}
 \end{equation}   
 Of course, we need to emphasize again  that here $t$ and $x$ should not be identified as time and space coordinate; they are just the coherent state expectation values, as given in (\ref{A4}). \\
 Since the physical states satisfying the Schr\"{o}dinger's equation are normalizable with respect to the inner product (\ref{op1}) of $L_{\star}^2(\mathbb{R}^1)$, it clearly indicates the fact that the physical states $\psi_{phy}(\hat{x},\hat{t})$ should belong to a suitable subspace  of $\hat{\mathcal{A}}_{\theta}$ (\ref{e59}) which is distinct from $\mathcal{H}_q$, for which the norm is obtained from the inner product defined for $L_{\star}^2(\mathbb{R}^2)$ (\ref{innpro_voros}).
\section{Forced harmonic oscillator in quantum space-time}
Having developed the basic formalism, we are now ready to discuss some application. Forced harmonic oscillator \cite{mehta,caru} is a good and simple example to study the QM of time dependent system as it has various applications in physics specially in quantum optics, as mentioned earlier. To this end, we place a forced harmonic oscillator in a quantum space-time in 1+1 dimension to investigate whether any effect of non-commutativity shows up in the geometric phase (which is not present in the system in classical space-time setting), when the system is evolved adiabatically in a closed loop in parameter space for a time period $\mathcal{T}$.\\ 
The Hamiltonian of a 1 D forced Harmonic oscillator (FHO) in  commutative space-time is given by,
\begin{equation}
H=\frac{\hat{p}_q^2}{2m}+\frac{1}{2}m\omega^2\hat{q}^2+f(t)\hat{q}+g(t)\hat{p}_q\label{e6}
\end{equation}
where $\hat{p}_q$ and $\hat{q}$ satisfies the usual Heisenberg algebra : $[\hat{q},\hat{p}_q]=i$ and $(f(t)), g(t))$ are a pair of suitably chosen periodic functions of time with time period $\mathcal{T}$ to be specified later. \\
Now in presence of non-commutativity in space-time we take the Hamiltonian to be given by the following simple operator ordering prescription, to render it hermitian:
\begin{equation}
\hat{H}=\frac{\hat{P}_x^2}{2m}+\frac{1}{2}m\omega^2\hat{X}^2+\frac{1}{2}[f(\hat{T})\hat{X}+\hat{X}f(\hat{T})]+g(\hat{T})\hat{P}_x\label{e7}
\end{equation}
where the coordinate space variables are replaced with $\hat{T}$ and $\hat{X}$ given by (\ref{e5}). Using this Hamiltonian in the Schr\"{o}dinger constraint equation (\ref{e74}), and taking its overlap with the coherent state basis (\ref{e45}), we get by using (\ref{e5}),
\begin{align}
&i\partial_t (x,t|\psi_{phy})=(x,t|\hat{H}|\psi_{phy})\nonumber\\
\Rightarrow &i\partial_t \psi_{phy}(x,t)=\left[\frac{P_x^2}{2m}+\frac{1}{2}m\omega^2 X_{\theta}^2+\frac{1}{2}\{f(T_{\theta})X_{\theta}+X_{\theta}f(T_{\theta})\}+g(T_{\theta})P_x\right]\psi_{phy}(x,t)\label{e8}
\end{align}\footnote{
The hermiticity of the  Hamiltonian in the right hand side of (\ref{e8}) can be established as already discussed in section 3.0.1 after (\ref{48}).}
where we have made use of the coherent state representation of the phase space variables, which are given by (\ref{47},\ref{48},\ref{25}) \cite{partha},
\begin{equation*}
P_t=-i\partial_t;\,\,\,
P_x=-i\partial_x;\,\,\,
X_{\theta}= x+\frac{\theta}{2}(\partial_x-i\partial_t);\,\,\,
T_{\theta}=t+\frac{\theta}{2}(\partial_t+i\partial_x)
\end{equation*}
Interestingly, this $X_{\theta}$ and $T_{\theta}$ can be related to commutative $x$ and $t$, defined in (\ref{A4}), by making use of similarity transformations as,
\begin{equation}
	X_{\theta}= SxS^{-1};\,\,\,\,\,T_{\theta}=S^{\dagger}t(S^{\dagger})^{-1}\label{M4}
\end{equation}  
where 
\begin{equation}
S=e^{\frac{\theta}{4}(\partial_t^2+\partial_x^2)}e^{-\frac{i\theta}{2}\partial_t\partial_x}\label{M5}
\end{equation}
is a non-unitary operator and this operator can be used to define a map from $L^2_*(\mathbb{R}^1)$ to $L^2(\mathbb{R}^1)$ as, 
\begin{align}
S^{-1}\,:\,\,L_*^2(\mathbb{R}^1)\,\,&\to\,\,L^2(\mathbb{R}^1)\nonumber\\
S^{-1}\big(\psi_{phy}(x,t)\big)\, &:= \, \psi_c(x,t)\,\,\in\,\, L^2(\mathbb{R}^1)\label{M6}
\end{align} 
where we have used the notation $\psi_c$ to denote the elements of $L^2(\mathbb{R}^1)$ space. Now one can easily verify that the inner product (\ref{op1}) in $L^2_*(\mathbb{R}^1)$ becomes identical to its commutative counterpart in $L^2(\mathbb{R}^1)$ (see Appendix-A.3)
\begin{equation}
\Big \langle \psi_{phy}\, ,\, \phi_{phy} \Big\rangle_{*,\,t}= \langle\psi_c\,\,,\phi_c\rangle _t\,\,\,\,\forall\,\psi_{phy},\phi_{phy}\,\in\, L^2_*(\mathbb{R}^1)\label{M7}
\end{equation}
where we have made use of integration by parts and dropped some boundary terms. This equality shows that once in $L^2(\mathbb{R}^1)$, we can indeed replace non-local Voros star product, \textit{only} within the integral, with the local point-wise multiplication as in the usual commutative QM. It should be emphasised here that, although the results of the integration as a whole are equal in both sides of (\ref{M7}), the integrands, by themselves are not : $\rho_{\theta}(x,t)=\psi_{phy}^*(x,t)\star \psi_{phy}(x,t)\,\ne\,\psi_c^*(x,t)\psi_c(x,t)=|\psi_c(x,t)|^2$, as certain boundary terms had to be dropped in deriving (\ref{M7}), as we have mentioned already. Physically, this means that both $\rho_{\theta}(x,t)$ and $|\psi_c(x,t)|^2$ can't be interpreted as the probability density simultaneously for observing the particle at the point $x$. And, as we have shown that $\rho_{\theta}$ (\ref{e75}) represents the probability density in NC spacetime, while $|\psi_c(x,t)|^2$ can be interpreted as probability density only in the commutative limit $\theta \to 0$. Further the integrands themselves are elements of two different algebras: non-commutative and commutative, which are not clearly $\star$-isomorphic \cite{bal1} to each other and the effect of non-commutativity is manifested in different fashion \cite{sg2} through the considered dynamical model when we go from $L_*^2(\mathbb{R}^1)$ to $L^2(\mathbb{R}^1)$ with help of the non-unitary transformation $S^{-1}$. \\
Using (\ref{e8},\ref{M4},\ref{M6}) ,can now be recast in the following form,
\begin{equation}
i\partial_t \psi_c(x,t)= \left[\frac{p_x^2}{2m}+\frac{1}{2}m\omega^2x^2+\frac{1}{2}\left\{f(S^{-1}S^{\dagger}t(S^{\dagger})^{-1}S)x+xf(S^{-1}S^{\dagger}t(S^{\dagger})^{-1}S)\right\}+g(S^{-1}S^{\dagger}t(S^{\dagger})^{-1}S)p_x\right]\psi_c(x,t)\label{e10}
\end{equation}
 Further using the identity $S^{-1}S^{\dagger}=e^{i\theta\partial_t\partial_x}:=U$, with $U$ being unitary, (\ref{e10}) can be simplified as,
\begin{equation}
i\partial_t \psi_c(x,t)=\left[\frac{p_x^2}{2m}+\frac{1}{2}m\omega^2x^2+\frac{1}{2}\{f(UtU^{-1})x+xf(UtU^{-1})\}+g(UtU^{-1})p_x\right]\psi_c(x,t) \label{e11}
\end{equation}
Now using the relation $UtU^{-1}=t-\theta p_x$, as follows by using Hadamard identity and the Taylor expansion up to first order in $\theta$, by taking $\theta$ to be very small, we can write,
\begin{equation}
	F(t-\theta p_x)\approxeq F(t)-\theta p_x \dot{F}(t)\label{e12}
\end{equation}
where $F(t)$ stands collectively for $f(t)$ and $g(t)$ and $\dot{F}(t)$ is their first order time derivatives. Note that $\dot{f}(t)$ and $\dot{g}(t)$ are also periodic functions of \textquoteleft{$t$}' like $f(t)$ and $g(t)$. We now impose the condition that $\dot{f}(t)$ and $\dot{g}(t)$ are slowly varying function of time. This will facilitate the use of adiabatic approximation in the system under consideration. Here we want to stress upon the fact that, had we taken $f(t)$ and $g(t)$ themselves to be adiabatic in nature, then second and higher order derivative of $f$ and $g$ would have to be dropped in our calculation because of adiabatic approximation and the geometric phase would vanish in that case. That is why we will choose such periodic functions $f(t)$ and $g(t)$ very judiciously, so that only their first order derivatives are slowly varying periodic functions of time.\\
Finally using (\ref{e12}) in (\ref{e11}) we get the effective commutative time dependent Schr\"{o}dinger equation for the theory as,
\begin{equation}
i\partial_t \psi_c(x,t)=H_c\psi_c(x,t)\label{e13}
\end{equation}
where $H_c$ is the corresponding effective commutative Hamiltonian and is given as
\begin{equation}
H_c= \alpha(t) p_x^2+\beta x^2+\gamma(t)(xp_x+p_xx)+f(t)x+g(t)p_x=H_{GHO}+ f(t)x+g(t)p_x\label{H}
\end{equation}
where $H_{GHO}$ stands for Hamiltonian of a generalised time dependent harmonic oscillator and the time dependent coefficients occurring there are given by,
\begin{equation}
\alpha(t)=\frac{1}{2m}-\theta \dot{g}(t);\,\,\,\,
\beta= \frac{1}{2}m\omega^2;\,\,\,\,
\gamma(t)=-\frac{1}{2}\theta\dot{f}(t)\label{e14}
\end{equation}
whereas the other two terms represent perturbations linear in position and momentum in coordinate basis. In order to diagonalize the whole Hamiltonian we try to diagonalize $H_{GHO}$ \cite{bdr} first. And for that we introduce annihilation and creation operators as follows:
\begin{equation}
a(t)=A(t)[x+(B(t)+iC(t))p_x]\label{e15}
\end{equation}
where,
\begin{align}
&A(t)= \sqrt{\frac{\beta}{\Omega (t)}}	\,\,\,\,\,\, \textrm{where}\,\,\,\,\,\Omega(t) = 2\sqrt{\alpha\beta - \gamma^2};\nonumber\\
&B(t)=\frac{\gamma(t)}{\beta};\,\,\,\,C(t)= \frac{\Omega(t)}{2\beta}\label{e17}
\end{align}
are time dependent real functions of time. Correspondingly,
\begin{equation}
a^{\dagger}(t)=  A(t)[x+(B(t)-iC(t))p_x]\label{e16}
\end{equation}
fulfilling $[a(t),a^{\dagger}(t)]=\textbf{1}$.\\
Note that the coefficients  $A, B , C$ and $\Omega$ are time dependent, as they depend upon $\alpha, \gamma$ and thus $\dot{f}(t)$ and $\dot{g}(t)$. So we can say that $A, B , C$ are slowly varying function of time and we shall in future neglect the second and higher order time derivative of these variables in our adiabatic approximation.\\
With this, the Hamiltonian (\ref{H}) can be written in terms of the ladder operator as , 
\begin{equation}
H_c= \Omega(t) \Big(a^{\dagger}(t)a(t)+\frac{1}{2}\Big)+P(t)a(t)+\bar{P}(t)a^{\dagger}(t)\label{H1}
\end{equation}
where, $$P(t)= A(t)[C(t)f(t)+i(B(t)f(t)-g(t))]=\sqrt{\frac{\beta}{\Omega(t)}}\left[\frac{\Omega}{2\beta}f(t)+i\left(\frac{\gamma(t)f(t)}{\beta}-g(t)\right)\right]$$
and $\Omega(t)$ can be identified with the \textquotedblleft{instantaneous} frequency". And although, this $\Omega(t)$ can be taken to be a slowly varying function of time, i.e. compatible with the adiabatic nature, the other time dependent functions  $P(t)$ and $\bar{P}(t)$ are a bit ambiguous as they consist of terms like the product of a fast varying and a slow varying functions. As a result, we cannot say that under time evolution the full Hamiltonian varies slowly.\\
In order to circumvent this problem let us now look for a suitable time dependent unitary transformation $\mathcal{U}(t)$, taking values in a suitable Lie group and transforming the wave function $\psi_c(x,t)$ in (\ref{e13}) as
\begin{equation}
\psi_c(x,t)\,\to\, \tilde{\psi}_c(x,t)=\mathcal{U}(t) \psi_c(x,t);\,\,\,\,\,\mathcal{U}^{\dagger}(t)\mathcal{U}(t)=1\label{M9}
\end{equation}
 With this, the corresponding Hamiltonian will transform under this time-dependent  unitary transformation as
 \begin{equation}
H_c\,\to\, \tilde{H}_c=\mathcal{U}(t)H_c\mathcal{U}^{\dagger}(t) - i\mathcal{U}(t)\partial_t \mathcal{U}^{\dagger}(t)\label{H3}
\end{equation}
This shows that the time evolution of the transformed states $\tilde{\psi}_c(x,t)$  is not governed by just $\mathcal{U}H_c\mathcal{U}^{\dagger}$ anymore, rather by an effective Hamiltonian $\tilde{\mathcal{H}}_c$, which is obtained by augmenting $\mathcal{U}H_c\mathcal{U}^{\dagger}$ by a suitable \textquotedblleft{connection}" like  term:
\begin{equation}
i\partial_t\tilde{\psi}_c(x,t)=\tilde{H}_c \tilde{\psi}_c(x,t) \label{H2}
\end{equation}
 The instantaneous energy eigenstates of the Hamiltonian $H_c$ are not anymore the eigenstates of those of $\tilde{H}_c$. Our purpose is to get rid of the terms involving fast variables like $P(t)$ and $\bar{P}(t)$ in $\tilde{\mathcal{H}}_c$ so that we are left with just the first term in (\ref{H1}) by a suitable choice of $\mathcal{U}(t)$. Indeed, it is quite gratifying to see that such a $\mathcal{U}(t)$ exists and is given by
 \begin{equation}
 	\mathcal{U}(t)= e^{-(wa-\bar{w}a^{\dagger}+il)},\,\,\,\,\,\,\,\,w\,\in \mathbb{C},\,\,l\,\in \,\mathbb{R}\label{e52}
 \end{equation} where
  $w$ and $l$ are given by,
 \begin{align}
 	w(t)&=-ie^{i\int_0^t\,dt'\,\Omega(t')}\int_0^t\,dt''\,P(t'') e^{-i\int_0^{t''}\,d\tilde{t}\,\Omega (\tilde{t}) } \nonumber\\
 	l(t)&= \int_0^t\,dt'\, \left[\Omega(t') |w(t')|^2-\frac{i}{2}\Big(w(t')\dot{\bar{w}}(t')-\dot{w}(t')\bar{w}(t')\Big)\right]\label{e18}
 \end{align}
 This structure of $\mathcal{U}(t)$ in (\ref{e52}) can be identified with the non-abelian Heisenberg group, obtained by exponentiating the elements of Heisenberg Lie algebra  $\textswab{h}$, generated by $(a,a^{\dagger},\mathbb{I})$, where $\mathbb{I}$ gives the central extension. 
With this we can show that (\ref{H3}) becomes,
\begin{equation}
\tilde{H}_c=\Omega(t)(a^{\dagger}a+\frac{1}{2})= H_{GHO}\label{H4}
\end{equation}
So by giving a time dependent unitary transformation (\ref{e52})  we are able to show that the total Hamiltonian $H_c$ (\ref{H}) transforms to a generalised hamonic oscillator Hamiltonian (\ref{H4}). Having diagonalized $\tilde{H}_c$, the computation of geometric phase acquired by $\psi_c(x,t)$ is reduced to that of obtaining the corresponding phase acquired by $\tilde{\psi}_c(x,t)$, which we take up in the next section.   
\section{Evolution of the ladder operators and appearance of geometric phase:}
It is quite clear that the time evolved form of $\tilde{\psi}_c(x,t)$ in (\ref{H2}) is considerably simpler to obtain than its precursor wave function $\psi_c(x,t)$  occurring in (\ref{e13}) and evolving with Hamiltonian $H_c$ (\ref{H}), using the diagonalised form of the corresponding Hamiltonian $\tilde{H}_c$  (\ref{H4}) in our adiabatic approximation, particularly because of the absence of terms involving the fast variables $P(t)$ and $\bar{P}(t)$. In fact, it becomes even more simpler to isolate the geometric phase in the Heisenberg picture, as was done in \cite{partha}. We undertake this task in this section.\\
 The Heisenberg equation of motion using (\ref{H4}) for the operators $a$ and $a^{\dagger}$ are given by,
\begin{equation}
\frac{da^{\dagger}}{dt}=\xi(t)a^{\dagger}+X(t)a\hspace{20pt}\textrm{and}\hspace{20pt} \frac{da}{dt}=\bar{\xi}(t)a+\bar{X}(t)a^{\dagger} \label{e20}
\end{equation}
where 
\begin{equation}
X(t)=-A^2(\dot{C}+i\dot{B});\,\,\,\,
\xi(t)=Y(t)+i\Omega(t);\,\,\,\,
Y(t)= A[(2\dot{A}C+A\dot{C})+iA\dot{B}]\label{e22}
\end{equation}
We now make use of the adiabaticity of the parameters $\alpha$ and $\gamma$ and thus those of $A,B,C$ by dropping their second and higher order time derivatives. The final form of the creation operator at the end of a cycle of time period $t=\mathcal{T}$ is given by (the detail of the calculation is provided in the Appendix-A.4),
\begin{equation}
a^{\dagger}(\mathcal{T})=a^{\dagger}(0)exp \left[i \int _0^{\mathcal{T}} \Omega\,\, d\tau + i\int_0^{\mathcal{T}} \left(\frac{1}{\Omega}\right)\frac{d\gamma}{d\tau} d\tau \right]\label{e35}
\end{equation}
The second term in the exponential of (\ref{e35}) represents the additional phase  over and above the dynamical phase $e^{i\int \Omega(t)dt}$, which is due the adiabatic evolution of the Hamiltonian around a closed loop $\Gamma$ in the parameter space in time $\mathcal{T}$. This geometric phase commonly known as Berry phase (in Heisenberg picture) can be written in a more familiar form, given as a functional of $\Gamma$,
  \begin{equation}
  	\Phi_G[\Gamma] =\oint_{\Gamma=\partial S} \frac{1}{\Omega} \nabla_\textbf{R}\gamma . d\textbf{R}=-\frac{\theta}{2}\int\int_S \nabla_{\textbf{R}}\left(\frac{1}{\Omega}\right) \times \nabla_{\textbf{R}} \Big(\dot{f}(t)\Big)\, .\, d\textbf{S}
  \label{e36}
  \end{equation}
   where we have used $\frac{d}{d\tau}= \frac{d\textbf{R}}{d\tau}\,.\,\nabla_\textbf{R}$, $\textbf{R}\equiv (\alpha,\beta,\gamma)$ being a vector in the parameter space whose components are time dependent and in the second equality, we have made use of Stoke's theorem to convert the line integral into the surface integral. The expression of the above equation (\ref{e36}), explicitly depends on the NC parameter $\theta$ and vanishes in the commutative limit $\theta \to 0$. Note that, we can relate this geometric phase shift obtained in the Heisenberg picture, to a more familiar form of Berry phase obtained by the state vectors, by going over to Schr\"{o}dinger picture.\\
  Finally observe that, had we worked in the Schr\"{o}dinger's picture, we would have obtained the same geometric phase $\Phi_G[\Gamma]$ acquired by the wave function $\tilde{\psi}_c(x,t)$ (\ref{M9}) in time $\mathcal{T}$. Eventually since the original wave function $\psi_c(x,t)$ is simply $\mathcal{U}^{\dagger}(t)\tilde{\psi}_c(x,t)$ with $\mathcal{U}(t)$ given in (\ref{e52},\ref{e18}) is a linear and unitary operator, it is clear that $\psi_c(x,t)$ too will acquire the same Berry phase in time $\mathcal{T}$, as the Berry phase, being a simple number (\ref{e36}) will not be affected by action of the Heisenberg group. We would like to make some pertinent observations before we conclude this section.\\ 
  \begin{itemize}
 \item{ It can be observed easily from (\ref{e36}) that the geometric phase depends upon the parameter $\gamma$ given in (\ref{e14}) which vanishes for $\theta = 0$. So the non-commutativity of the space-time plays here a crucial role for the appearance of the geometric phase. Note in this context that the authors in \cite{bdr} have shown that the occurrence of the dialatation term in the Hamiltonian is a necessary element for a system to produce the geometric phase. And although our original system i.e. the forced harmonic oscillator (\ref{e6}) did not contain such term, placing the system in a NC space-time has enabled us to generate such a term in the effective commutative Hamiltonian (\ref{H}).}
 \item{The Hamiltonian of the F.H.O in (\ref{H}) can be expanded in terms of the generators of the double photon algebra \cite{dasgupta}. For that let us define
  \begin{equation}
  K_+=\frac{ix^2}{2},\,\,\,K_-=\frac{ip^2}{2},\,\,\,K_0=\frac{i(xp+px)}{4},\,\,\, A_+=x, A_-=p\label{e39}
  \end{equation}
  Taking $\mathbb{I}$ as a central extension we can show that the above generators follow the following Lie algebra:
  \begin{align}
  	&[K_0,K_{\pm}]=\pm  K_{\pm},\,\,\,\,[K_+,K_-]=-2K_0\label{e40}\\
  	&[K_+,A_-]=-A_+,\,\,\,[K_-,A_+]=A_-,\,\,\,[K_0,A_+]=\frac{A_+}{2},\,\,\,[K_0,A_-]=-\frac{A_-}{2},\,\,\,[A_+,A_-]=i\mathbb{I}\label{e41}
  \end{align}  
  where the sub-algebra (\ref{e40}) is the su(1,1) Lie-algebra and (\ref{e40}),(\ref{e41}) as a whole represents  the double photon algebra \cite{dasgupta}.
 \\Now the Hamiltonian (\ref{H}) can therefore be written in terms of the generators (\ref{e39}) as
 \begin{equation}
 	H_c= B^{\mu}(t)K_{\mu}+f(t)A_++g(t)A_-;\,\,\,\,\mu=+,-,0;\,\,\,\,B^+=-2i\alpha(t),B^-=-2i\beta(t), B^0=-2i\gamma(t)\label{e42}
 \end{equation} 
 The corresponding transformed Hamiltonian $\tilde{H}_c$, on the other hand, takes its value only in the su(1,1) sub-algebra.
 \begin{equation}
 	H_c \,\to \tilde{H}_c\, = \mathcal{U}H_c\mathcal{U}^{\dagger}-i\mathcal{U}\frac{\partial \mathcal{U}^{\dagger}}{\partial t}= B^{\mu}(t)K_{\mu}
 \end{equation}  
At this stage, we may point out that the Hamiltonians in \cite{bdr} were also, for example, su(1,1) and su(2) Lie algebra valued, in analogy with ours. So like in their case, we too can anticipate the occurance of geometric phase in our case.}
\end{itemize}
 \section{Conclusion and future direction}
  Experimental detection of Planck scale phenomena involving quantum structure of space-time, in a non-relativistic (NR) system, is a challenging and difficult task. However, in some previous works in the literature \cite{cast,bek}, authors have tried to find plausible scenarios, which can possibly help to find out the effects caused by quantum structure of space-time, in low energy regime. That motivated us to study the fingerprinters of Planck scale physics in NR quantum system, which is the main objective of our present work. Here we will briefly summerize our findings.\\
 We have shown that a system of a forced harmonic oscillator, when placed in a non-commutative space-time background, acquires a geometric phase shift. The effective commutative system Hamiltonian is shown to be as that
 of a time dependent generalised harmonic oscillator with perturbation linear in position and momentum. The system is then subjected to a unitary transformation to get rid of these perturbations and then diagonalised with the help of the time dependent unitary transformation. We then derived the equation of motion of the ladder operators in Heisenberg picture, which shows that an extra phase over and above the dynamical phase is produced, depending on the geometry of the parameter space, when the system is transported adiabatically around a closed loop. This phase directly depends on the non-commutative parameter $\theta$ in our expression, computed upto first order in $\theta$ and vanishes in commutative limit $\theta \to 0$. It should be noted in this context that our computation, carried out in Heisenberg picture, as in \cite{pnskp}, is rather a bit of an unconventional approach to calculate the geometric phase. Its primary advantage is that, it provides a natural framework for semi-classical
 correspondence. The geometric phase calculated in this procedure can be shown to be equal to the Hannay's angle, which is generated by the corresponding classical adiabatic evolution.\\ \\
 Finally, we would like to comment on some future prospects of our work.
 In \cite{anan}, the authors had shown an interesting relationship between the Fubini-Study metric defined in the projective Hilbert space of a time dependent quantum system and the energy-time uncertainty relation. Later,
 authors in \cite{skb} have pointed out that, the coherent states provide a natural framework for the realization of the above mentioned relationship and demonstrated this explicitly  in their computation involving a spin system in a magnetic field and a generalised harmonic oscillator system respectively. In light of the present paper, it seems that we can also extend our study to compute the energy-time uncertainty relation in our system as well, by making use of the existing Fubini-Study metric of the projective Hilbert space. Presumably, the space-time noncommutativity, which is solely responsible for giving rise to the geometric phase and inducing a geometry in the parameter space, should have some tangible effect in the energy-time uncertainty relation and it should be quite interesting to investigate this point in our future work.\\ 
 On the other hand, in \cite{mehta}, it was shown that a coherent state persists to be coherent at all times if the system Hamiltonian, with which the operators are evolved (in Heisenberg picture), is that of a time dependent forced harmonic oscillator. We can extend this computation for our system, which is basically FHO, in the presence of noncommutativity in space-time. The effective commutative system, which is a system of generalised forced harmonic oscillator (GFHO), will produce states which are squeezed coherent states of a usual FHO \cite{xu}. This encourages us to study the fate of these squeezed coherent states under time evolution and its application in quantum optics and quantum information theory, if any. \\ \\
Furthermore, note that the idea of noncommutativity in the space-time sector may also be introduced through the non-relativistic,
second quantized formalism. This would indeed be a good starting point for considering NC space-time because, unlike in
QM, here, space and time can naturally be treated on an equal footing, in
the sense that the spatial coordinates do not correspond to any observables anymore and to begin with, they are just c-numbered parameters (labelling the continuous spatial degrees of freedom of the system) like the evolution parameter time. Importantly, the second quantized formalism allows us to investigate the impact of space-time noncommutativity on the one-particle sector of quantum
field theory for a generic external potential \cite{nair}, which is similar to NC quantum mechanical analysis in the first quantized formalism. We believe that the nontrivial changes in the Schr\"{o}dinger equation are indeed originating
from the space-time noncommutativity which has not hitherto been explored in the literature, to the best of our knowledge. We hope to return with
some of these issues in our future works.
\section*{Acknowledgement}
P.N. would like to thank Prof. A. Deriglazov for valuable and constructive suggestions during the development of the work and Prof. A.P. Balachandran for carefully going through the manuscript and pointing out certain shortcomings in the earlier version of the draft. A.C. would like to express sincere gratitude to Mr. Sayan Kumar Pal for many helpful discussions for this paper. A.C. would like to thank DST-India for providing financial support in the form of fellowship during the course of this work and P.N. would like to thank S N Bose National Centre for Basic Sciences, Kolkata, for providing financial support during the project tenure. 
\section*{Appendices}
\subsection*{A.1\,\,\,\,On the effect of space-time noncommutativity on a time-independent system}
 Let us take a time-independent Hamiltonian placed in a NC space-time.   
 \begin{equation}
 H= \frac{\hat{p}_x^2}{2m}+V(\hat{X})
 \end{equation}
 The corresponding Schr\"{o}dinger equation is obtained by taking its representation in coherent state basis (\ref{A8}) to get,
\begin{equation}
 i\partial_t\psi_{phy}(x,t)=(x,t| \hat{H}|\psi_{phy}) =\Big[\frac{p_x^2}{2m}+V(X_{\theta})\Big]\psi_{phy}(x,t);\qquad \psi_{phy}(x,t) \in L^2_*(\mathbb{R}^1)
 \end{equation}
 Now, using (\ref{M4}), we get
 \begin{equation*}
 i\partial_t \psi_{phy}(x,t)= \Big[\frac{p_x^2}{2m}+ SV(x)S^{-1}\Big]\psi_{phy}(x,t)
 \end{equation*}
 which, when re-written in terms of $\psi_c(x,t)=S^{-1}\psi_{phy}(x,t)\in L^2(\mathbb{R}^1)$ (\ref{M6}), readily yields the usual commutative Schr\"{o}dinger equation:
 \begin{equation}
 i\partial_t\psi_c(x,t) = \big[\frac{p_x^2}{2m}+V(x)\Big] \psi_c(x,t)\label{N85}
 \end{equation}
thereby showing that one effectively recovers the commutative Schr\"{o}dinger equation and leaves the spectrum and therefore the dynamics of the system unaffected. Rather, it is the wave function of the system that only gets modified by the transformation : $\psi_c=S^{-1}\psi_{phy}$.
\subsection*{A.2\,\,\,Dirac's constraint analysis for the Lagrangian (\ref{e73})}
Starting with the first order form of the Lagrangian (\ref{e73}), here we are going to carry out Dirac's constraint analysis and compute the Dirac's bracket among the configuration space variables where space-time and their corresponding canonical momenta are considered to be configuration space variable,
\begin{equation}
L_f^{\tau,\theta}= p_{\mu}\dot{x}^{\mu}-\sigma(\tau)(p_t+H)+\frac{\theta}{2}\epsilon^{\mu\nu}p_{\mu}\dot{p}_{\nu},\,\,\,\,\,\,\mu,\nu=0,1 \label{e80}
\end{equation}
$\sigma(\tau)$  is an arbitrary Lagrange multiplier. Now we can derive the canonical momenta corresponding to $x^{\mu}, p_{\mu}$ and $\sigma(\tau)$ as follows :
\begin{equation}
\pi^x_{\mu}=\frac{\partial L^{\tau,\theta}}{\partial \dot{x}^{\mu}}=p_{\mu};\qquad
\pi_p^{\mu}=\frac{\partial L^{\tau,\theta}}{\partial \dot{p}_{\mu}}=-\frac{\theta}{2}\epsilon^{\mu\nu}p_{\nu}; \qquad
\pi_{\sigma}=\frac{\partial L^{\tau,\theta}}{\partial\dot{\sigma}}=0\label{e81}
\end{equation}
We can see that the canonical momenta are not related to the generalized velocities and so that the above equations (\ref{e81}) are interpreted as primary constraints of the theory and can be written as

\begin{equation}
\Phi_{1,{\mu}}=\pi_{\mu}^x-p_{\mu}\approx 0,\,\,\,\,\,\,\,\Phi_2^{\mu}=\pi^{\mu}_p+\frac{\theta}{2}\epsilon^{\mu\nu}p_{\nu}\approx 0,\,\,\,\,\,\,\,\Phi_3 = \pi_\sigma \approx 0\label{e82}
\end{equation}
Using the following Poission brackets between the canonical pairs,
\begin{equation}
\{x^{\mu},\pi_{\nu}^x\}=\delta^{\mu}\,_{\nu},\,\,\,\,\,\,\{p_{\mu},\pi^{\nu}_p \} =\delta_{\mu}\,^{\nu}
\end{equation}
 we can derive the Poission brackets between the constraints as (\ref{e82}) as
\begin{align}
&\{\Phi_3,\Phi_3\} =\{\Phi_3,\Phi_{1,\mu}\} = \{\Phi_3,\Phi_2^{\mu}\} =0\nonumber\\
&\{\Phi_{1,{\mu}},\Phi_{1,{\nu}}\} =0;\qquad
\{\Phi_{1,{\mu}},\Phi_2^{\nu}\} = -\delta_{\mu}\,^{\nu}\qquad
\{\Phi_2^{\mu},\Phi_2^{\nu}\}= \theta\epsilon^{\mu\nu}\label{e84}
\end{align}
We can therefore write the canonical Hamiltonian as
\begin{equation}
H_c= \sigma(\tau)(p_t+H)\label{e85}
\end{equation}
We can see that $\Phi_3$ gives zero brackets with all other constraints. So it can be classified as a first class constraint. The rest of the constraints are classified as second class constraints as the Poission bracket between $\Phi_{1,{\mu}}$ and $\Phi_{2,\mu}$ does not vanish. We can therefore implement the second class constraints strongly through the Dirac brackets. But, before that note that, the total Hamiltonian can be written by adding all the constraints to the canonical Hamiltonian $H_c$ (\ref{e85}) as,
\begin{equation}
H_T=\sigma(\tau)(p_t+H)+\lambda_{1,\mu}\Phi_{1,\mu}+\lambda_{2,\mu}\Phi_2^{\mu}+\lambda_3\Phi_3\label{e86}
\end{equation}
where $\lambda_{1,\mu},\lambda_{2,\mu}$ and $\lambda_3$ are suitable multipliers. Now the preservation of $\Phi_3$ in time yields the secondary constraint:  
\begin{equation}
\Sigma=\dot{\Phi_3}=\{H_T,\pi_{\sigma}\} = p_t+H \approx 0 \label{e87}
\end{equation}
From the time preservation conditions of the other constraints we just get the relation between $\sigma(\tau)$ and $\lambda_{1,\mu},\lambda_{2,\mu}$, which are not of our concern in this context. Rather, we shall implement the other second class constraints, as mentioned above. And for that we first need to write down the constraint matrix $\Lambda_{ab}$ and its inverse $(\Lambda^{-1})_{ab}$ which are now obtained as,
\begin{equation}
\Lambda_{ab}= \begin{pmatrix}
\{\Phi_{1,{\mu}},\Phi_{1,{\nu}}\}&\{\Phi_{1,{\mu}},\Phi_2^{\nu}\}\\
\{\Phi_2^{\nu},\Phi_{1,{\mu}}\}&\{\Phi_2^{\mu},\Phi_2^{\nu}\}
\end{pmatrix}=\begin{pmatrix}
0&-\delta_{\mu}\,^{\nu}\\
\delta^{\nu}\,_{\mu}&\theta\epsilon^{\mu\nu}
\end{pmatrix};\,\,\,\,\,\,\,\,\,
(\Lambda^{-1})_{ab}= \begin{pmatrix}
\theta \epsilon_{\mu\nu}&\delta^{\mu}\,_{\nu}\\
-\delta_{\nu}\,^{\mu}&0
\end{pmatrix}\label{e89}
\end{equation}
fulfilling, $\Lambda_{ab}(\Lambda^{-1})_{bc}=\delta_{ac}$.\\
Using the following definition for Dirac brackets \cite{dir},
\begin{equation}
\{f,g\}_D= \{f,g\}-\{f,\Phi_a\}(\Lambda^{-1})_{ab}\{\Phi_b,g\}
\end{equation}
we now compute the Dirac brackets between the phase space variables to get the following structure (\ref{lev}):
\begin{equation}
\{x^{\mu},x^{\nu}\}_D=\theta\epsilon^{\mu\nu},\,\,\,\,\,\,
\{p_{\mu},p_{\nu}\}_D=0,\,\,\,\,\,\,
\{x^{\mu},p_{\nu}\}_D=\delta^{\mu}\,_{\nu}\label{A0}
\end{equation}
These Dirac brackets, when elevated to the quantum level, readily yield the NCHA (\ref{e1},\ref{e2}), which includes the space time noncommutative structures.\\
Finally, we would like to make certain observations regarding the roles of primary first class constraint $\Phi_3$ and the secondary constraint $\Sigma$. The primary first class constraint $\Phi_3$ (\ref{e82}) involves the canonical momentum corresponding to the Lagrange's multiplier $\sigma(\tau)$, which is not physical in nature. So we can ignore this constraint. On the other hand by using the Dirac brackets (\ref{A0}) we can easily show that the secondary constraint $\Sigma$ (\ref{e87}) is a first class constraint as it has vanishing Dirac brackets with all other primary constraints. And eventually $\Sigma$ can be shown to give rise to the $\tau$ evolution of the system in the form of gauge transformation of the theory.
\subsection*{A.3\,\,\, On the mapping $S^{-1}$: $L_*^2(\mathbb{R}^1)$ to $L^2(\mathbb{R}^1)$}
Here we would like to prove that, the map (\ref{M6}) i.e. $S^{-1}:\,L^2_*(\mathbb{R}^1)\,\to\,L^2(\mathbb{R}^1)$, with $S^{-1}=e^{-\frac{\theta}{4}(\partial_t^2+\partial_x^2)}e^{i\frac{\theta}{2}\partial_t\partial_x}$, a non-unitary operator, is an inner product preserving one i.e. the inner product between a pair of arbitrary states in $L_*^2(\mathbb{R}^1)$ space with respect to star multiplication, is equal to the inner product of the transformed states belonging to $L^2(\mathbb{R}^1)$ with respect to ordinary point-wise multiplication.\\
For that, let $\psi_{phy}(x,t;\theta)\,\in\,L^2_*(\mathbb{R}^1)$ and $\psi_c(x,t)\,\in\,L^2(\mathbb{R}^1)$ and related by (\ref{M6}) as $\psi_{phy}(x,t,\theta)=S\psi_c(x,t)$, satisfy the respective Schr\"{o}dinger equations (\ref{e8}) and (\ref{e13}).
Now, if we consider a stationary wave-function in $L^{2}(\mathbb{R}^1)$ as 
\begin{equation}
\psi_c(x,t)= e^{-iEt}\psi_c(x),    \label{4}
\end{equation}
then the corresponding physical state in $L^{2}_{\star}(R^{1})$ can be easily determined by using (\ref{M6}) as
\begin{equation}
\psi_{phy}(x,t;\theta)= e^{-iEt} e^{\frac{\theta}{4}(-E^{2}+\partial^{2}_{x})}e^{\frac{\theta}{2}E\partial_{x}} \psi_c(x)=e^{-iEt}\psi_{phy}(x;\theta)\label{1}
\end{equation}
So here too we see that, the time part gets factored out here in the form of $e^{-iEt}$ just like (\ref{4}), which is indicative of the fact that this $\psi_{phy}(x,t,\theta)$ is also a stationary state with the same energy eigenvalue $E$. Now the norm of the state $\psi_c(x,t)\in L^{2}(\mathbb{R}^1)$ is defined as 
\begin{align}
&\int\,\,dx\,\,\psi_c^*(x,t)\psi_c(x,t)\nonumber\\
=&\int\,\,dx\,\, (S^{-1}\psi_{phy}(x,t; \theta))^*\quad (S^{-1}\psi_{phy}(x,t; \theta))\nonumber\\
=&\int\,\,dx\,\,\Big[\psi^*_{phy}(x,t;\theta)e^{-\frac{\theta}{4}(-E^2+\overleftarrow{\partial_x}^2)}e^{-\frac{i\theta}{2}\overleftarrow{\partial_t}\overleftarrow{\partial_x}}\Big]\Big[e^{-\frac{\theta}{4}(-E^2+\overrightarrow{\partial_x}^2)}e^{i\frac{\theta}{2}\overrightarrow{\partial_t}\overrightarrow{\partial_x}}\psi_{phy}(x,t;\theta)\Big]\nonumber\\
=&\int\,\,dx\,\,\Big[\psi^*_{phy}(x,t;\theta)e^{-\frac{\theta}{4}(-\overleftarrow{\partial}_t\overrightarrow{\partial}_t-\overleftarrow{\partial_x}\overrightarrow{\partial}_x)}e^{\frac{i\theta}{2}\overleftarrow{\partial_t}\overrightarrow{\partial_x}}\Big]\Big[e^{-\frac{\theta}{4}(-\overleftarrow{\partial}_t\overrightarrow{\partial}_t-\overleftarrow{\partial_x}\overrightarrow{\partial_x})}e^{-i\frac{\theta}{2}\overleftarrow{\partial_x}\overrightarrow{\partial_t}}\psi_{phy}(x,t;\theta)\Big]\nonumber\\
&\textrm{(Here we have used integration by parts and dropped the boundary terms.)}\nonumber\\
=&\int\,\,dx\,\, \Big[\psi^*_{phy}(x,t;\theta)e^{\frac{\theta}{2}(\overleftarrow{\partial}_t\overrightarrow{\partial}_t+\overleftarrow{\partial}_x\overrightarrow{\partial}_x)}e^{\frac{i\theta}{2}(\overleftarrow{\partial}_t\overrightarrow{\partial}_x-\overleftarrow{\partial}_x\overrightarrow{\partial}_t)}\psi_{phy}(x,t;\theta)\Big]=\int\,\,dx\,\, \psi^*_{phy}(x,t;\theta)\star_V\psi_{phy}(x,t;\theta)\label{3}
\end{align}
With this we are able to prove that, the norm of the stationary wave function $\psi_{phy}(x,t;\theta)$ associated with energy eigen value $E$, in $L^2_*(\mathbb{R}^1)$ space is equal to the norm of the transformed stationary wave-function $\psi_c(x,t)$ in $L^2(\mathbb{R}^1)$ space with the same energy eigenvalue $E$. The products defining the norms in Hilbert spaces $L^2(\mathbb{R}^1)$ and $L^2_*(\mathbb{R}^1)$ are the usual point-wise multiplication and Voros star multiplication respectively.\\ \\
Now let us construct a pair of non-stationary physical states by taking linear combination of stationary state wave functions with suitable coefficients such as,
\begin{equation}
\Psi_{phy}(x,t;\theta)=\sum_n A_n \psi^{(n)}_{phy}(x,t;\theta), 
\label{Df}
\end{equation}
\begin{equation}
\Phi_{phy}(x,t;\theta)=\sum_{n} B_{n} \psi^{(n)}_{phy}(x,t;\theta) \label{2}
\end{equation}
where $\psi^{(n)}_{phy}(x,t;\theta)=e^{-iE_nt}\psi^{(n)}_{phy}(x;\theta)$ is a stationary state for a particular $n^{th}$ energy eigen value $E_n$. Note that in 1D spatial dimension all bound stationary states are non-degenerate and here  $\psi^{(n)}_{phy} (x,t;\theta)$ forms a complete set of basis and also therefore can be chosen to orthonormal as, 
\begin{equation}
\int_{t} ~dx ~\psi^{*(n)}_{phy}(x,t;\theta)\star_{V}  \psi^{(m)}_{phy}(x,t;\theta)=\delta_{mn}
\label{lk}
\end{equation}
Now let us consider the inner products  between this pair of  non stationary states (\ref{Df}) and (\ref{2})  as 
\begin{equation}
\int dx ~\Phi^{\star}_{phy}(x,t;\theta)\star_{V}\Psi_{phy}(x,t;\theta)=\sum_{n,m} A_{m}B^{\star}_{n} \int dx ~\psi^{(n) \star}_{phy}(x,t;\theta)\star_{V}\psi^{(m)}_{phy}(x,t;\theta)
\label{p}
\end{equation}
Using the orthonormality (\ref{lk}) we arrive at 
\begin{align}
\int\, dx \,\Phi^{\star}_{phy}(x,t;\theta)\star_{V}\Psi_{phy}(x,t;\theta)&=\sum_{n} A_{n}B^{\star}_{n} \int dx ~\psi^{(n) \star}_{phy}(x,t;\theta)\star_{V}\psi^{(n)}_{phy}(x,t;\theta)\nonumber\\
(\textrm{Now using (\ref{3})})\qquad &=\sum_n A_nB_n^* \,\int \,\,dx\,\,\psi_c^{(n)*}(x,t)\,\psi_c^{(n)}(x,t)\label{hj},
\end{align}
where we have used  $\psi_c^{(n)}(x,t)=S^{-1}\psi^{(n)}_{phy}(x,t;\theta) $.\\
It can be shown by simple argument that, if the non-stationary state $\Psi_{phy}(x,t,\theta)\,\in\,L^2_*(\mathbb{R}^1)$ is given by (\ref{Df}), then the corresponding non-stationary state in $L^2(\mathbb{R}^1)$ is given by, 
\begin{equation}
\Psi_c(x,t) =S^{-1}\Psi_{phy}(x,t,\theta)=S^{-1}\sum_n A_n \psi_{phy}^{(n)}(x,t,\theta)=\sum_n A_n S^{-1}\psi_{phy}^{(n)}(x,t,\theta)=\sum_n A_n \psi_c^{(n)}(x,t)
\end{equation}
Using this we can finally write (\ref{hj}) as,
\begin{equation}
\int\,\,dx\,\, \Phi^*_{phy}(x,t,\theta)\,\star_V\,\Psi_{phy}(x,t,\theta)= \int\,\,dx\,\, \Phi_c^*(x,t)\,.\,\Psi_c(x,t)    
\end{equation}
This equality suggests that, the  inner product between two non-stationary states in $L^2_*(\mathbb{R}^1)$ with respect to star multiplication is indeed equal to the inner product between the corresponding pair of non-stationary states of $L^2(\mathbb{R}^1)$ space with respect to usual point-wise product which we use for usual commutative QM. This is also expected because of the following reason: the map, $S^{-1}$ is a non-unitary transformation, which naturally changes the inner-product of the Hilbert space and the Voros star product in $L^2_*(\mathbb{R}^1)$ gets replaced by usual point-wise product of $L^2(\mathbb{R}^1)$ space.  
\subsection*{A.4\,\,\,Calculation of the geometric phase}
To solve the first order differential equation of the ladder operators given by (\ref{e20}), we can decouple the equations in  (\ref{e20}) , by taking their time derivatives and combining them suitably to get \cite{pnskp},
\begin{equation}
\frac{d^2a^{\dagger}}{dt^2}+Z_1(t)\frac{da^{\dagger}}{dt}+Z_2(t)a^{\dagger}=0;\,\,\,\,\,\,\,\,\,\frac{d^2a}{dt^2}+\bar{Z}_1(t)\frac{da}{dt}+\bar{Z}_2(t)a=0\label{e24}
\end{equation}
where 
\begin{equation}
Z_1(t)= -\left(\frac{\dot{X}}{X}+\xi+\bar{\xi}\right);\qquad
Z_2(t)=-(\dot{\xi}-\frac{\dot{X}}{X}\xi +X\bar{X}-\xi\bar{\xi})\label{e25}
\end{equation}
where $X(t), \xi(t)$ are time dependent and defined in (\ref{e22}). All the expressions found here, so far,  are exact. We shall now  consider the adiabaticity of $\alpha(t)$ and $\gamma(t)$ in (\ref{e14}). Since, we have taken  $\dot{f}$ and $\dot{g}$ to be slowly varying periodic function of time, it follows that, $\alpha(t)$ and $\gamma(t)$ (\ref{e14}) too vary adiabatically with time. Further, since $A(t),B(t)$ and $C(t)$ (\ref{e17}) depends linearly on $\alpha(t),\beta, \gamma(t)$, they too follow the same order of adiabaticity. So if $\dot{\alpha},\dot{\gamma} \sim \epsilon$ and $\ddot{\alpha},\ddot{\gamma}\sim \epsilon^2$, then, $\dot{F}\sim \epsilon\,\, \textrm{and}\,\, \ddot{F}\sim \epsilon^2$,where $F$ collectively stands for $A,B,C$. Eventually we shall ignore second and higher order time derivatives ($\sim \epsilon^2$)  for considering adiabatic time evolution. To proceed further we write (\ref{e24}) in its normal form. To do so, we introduce another time dependent operator, $b(t)$ as,
\begin{equation}
a^{\dagger}(t)= b^{\dagger}(t)e^{-\frac{1}{2}\int_0^t Z_1(\tau) d\tau}\label{e26}
\end{equation}
With this the first equation of (\ref{e24}) can be recast in terms of $b^{\dagger}$
\begin{equation}
\frac{d^2b^{\dagger}(t)}{dt^2}+\tilde{Z}_2(t)b^{\dagger}(t)=0\label{e27}\end{equation}
where, 
\begin{align}
&\tilde{Z}_2=Z_2-\frac{1}{2}\dot{Z}_1-\frac{Z_1^2}{4}\approx \Omega^2-\Omega \tilde{W}+i (2\Omega\frac{\dot{A}}{A}+W) \sim \mathcal{O}(\epsilon) = U+iV (say)\label{e29}
\end{align} 
where we have taken $\frac{\ddot{C}+i\ddot{B}}{\dot{C}+i\dot{B}}=W+i\tilde{W} \sim \mathcal{O}(\epsilon)$.
Then  (\ref{e27}) can be re-written in the following form: 
\begin{equation}
\frac{d^2b^{\dagger}(t)}{dt^2}+(U+iV)b^{\dagger}(t)=0\label{e28}
\end{equation}
As we are working in the adiabatic regime, both $U$ and $V$ vary slowly with time as is clear from (\ref{e29}). We can then apply the formula for WKB approximation for complex potential \cite{mohr} given by,
\begin{equation}
b^{\dagger}(t)=b^{\dagger}(0)\left[\frac{C_1}{\sqrt{|\chi(t)|}}exp\left(\int_{0}^{t}(i\chi(\tau)-\phi(\tau)) d\tau\right)+\frac{C_2}{\sqrt{|\chi(t)|}}exp\left(\int_{0}^{t}(-i\chi(\tau)+\phi(\tau))d\tau\right)\right]\label{e31}
\end{equation}
where $\sqrt{U+iV}=\chi+i\phi$ so that (\ref{e31}) can be applied  to find the general solution of the differential equation (\ref{e28}). In our case, $\chi$ and $\phi$ are given as,
$$\chi=\sqrt{\frac{\sqrt{U^2+V^2}+U}{2}}\approx \sqrt{U+\frac{V^2}{4U}}\approx \sqrt{U} \approx \Omega - \frac{\tilde{W}}{2}$$
$$\phi= \sqrt{\frac{\sqrt{U^2+V^2}-U}{2}}\approx \sqrt{\frac{V^2}{4U}}\approx \frac{\dot{A}}{A}+\frac{W}{2}$$
where second and higher order derivative terms are ignored everywhere. We also have implied the parameters to be periodic. Further note that only the first term in the exponential of (\ref{e31}) yields the correct sign of the dynamical phase for $a^{\dagger}$. So we shall put $C_2 =0$. Now combining all the expressions we get at $t=\mathcal{T}$,
\begin{equation}
b^{\dagger}(\mathcal{T})\approx b^{\dagger}(0)exp \left(\int_0^{\mathcal{T}}[i\Omega-\frac{d}{d\tau}(ln A)-\frac{1}{2}(W+i\tilde{W})] d\tau\right) \label{e33}
\end{equation}
As the integrands are exact differential and as we have the periodic boundary condition for the functions, i.e. $A(T)=A(0)$, we can drop some terms in the exponent as follows: 
$$\int_0^{\mathcal{T}}\Big(\frac{\dot{A}}{A}+ \frac{W+i\tilde{W}}{2}\Big) d\tau = \int_0^{\mathcal{T}} \frac{1}{2}\frac{d}{d\tau}\Big(ln X\Big) d\tau =0$$
$X$ is given in (\ref{e22}). Substituting (\ref{e33}) in (\ref{e26}), we get
\begin{equation}
a^{\dagger}(\mathcal{T})=b^{\dagger}(\mathcal{T})\,\,exp\,\,\Big({\int_0^{\mathcal{T}} \Big[\frac{d}{d\tau}(ln\,X+A^2C)+iA^2B\Big] d\tau}\Big)= b^{\dagger}(\mathcal{T})e^{i\int_0^{\mathcal{T}} A^2\dot{B} d\tau}\label{e32}
\end{equation}
The expression of $A,B,C$ are given in (\ref{e17}). Finally making use of  (\ref{e33}) in (\ref{e32}) we can ultimately reproduce (\ref{e35}) as,
\begin{align}
a^{\dagger}(\mathcal{T})&=a^{\dagger}(0)exp \left(i\int_0^{\mathcal{T}}\Omega\,\,d\tau \right)exp\left(i\int_0^{\mathcal{T}}A^2 \dot{B}\,\,d\tau \right)  \nonumber\\
&=a^{\dagger}(0)exp \left[i \int _0^{\mathcal{T}} \Omega\,\, d\tau + i\int_0^{\mathcal{T}} \left(\frac{1}{\Omega}\right)\frac{d\gamma}{d\tau}\,d\tau\,\,\, \right]\label{e34}
\end{align}
The two terms in the exponents clearly represent the dynamical and geometric phase respectively.


\begin{thebibliography}{9}
\bibitem{bron}\emph{M.P. Bronstein , Zh. Eksp. Teor. Fiz., \textbf{6} (1936).}
\bibitem{dop}\emph{S. Doplicher, K. Fredenhagen ,J. E. Roberts,Commun. Math. Phys. \textbf{172}, (1995)  p. 187-220.}
\bibitem{szabo} \emph{R. J. Szabo, Physics Reports \textbf{378} (2003) 207-299.}
\bibitem{apb}\emph{A. P. Balachandran, T. R. Govindarajan, A. G. Martins and P. Teotonio-Sobrinho, J. High Energy Phys.
\textbf{0411}, (2004) 068.\\
 A. P. Balachandran and A. Pinzul, Mod. Phys. Lett. A \textbf{20}, (2005) 2023.}
\bibitem{qft}\emph{M. Chaichian, A. Demichev, P. Presnajder, and A. Tureanu, Eur. Phys. J. C \textbf{20},(2001) 767.\\
 L. AlvarezGaume, J. L. F. Barbon, and R. Zwicky, J. High Energy Phys. \textbf{05}, (2001) 057.}
 \bibitem{liz}\emph{A. P. Balachandran and L. Chandar, Nucl. Phys. B \textbf{428} (1994)
 435.\\
 F Lizzi, P Vitale, Phys. Lett. B, \textbf{818} (2021) 136372.}
\bibitem{time}\emph{Paul Busch, Marian Grabowski, Pekka J. Lahti, Phys. Lett. A, \textbf{191} (1994) 357-361.}
\bibitem{hooft}\emph{G. ’t Hooft, Class. Quant. Grav. \textbf{10} (1993) 1653.}
\bibitem{bal2}\emph{A. P. Balachandran, A. Joseph, P. Padmanabhan, Phys. Rev. Lett.\textbf{105},(2010) 051601.\\
A.P. Balachandran, P. Padmanabhan, JHEP \textbf{12}(2010) 001.}
\bibitem{sg1}\emph{S. Ghosh, P. Pal, Phys. Lett. B, \textbf{618} (2005) 1–4, P.243-251.}
\bibitem{partha} \emph{P. Nandi, S. K. Pal, A. N. Bose, B. Chakraborty, Ann. Phys. \textbf{386} (2017) 305-326.}
\bibitem{pau}\emph{ W. Pauli, Handbook der Physics, edited by S. Flugge, Vol 5/1 (Berlin, 1926), p.60.}
\bibitem{v1}\emph{N. G. Sanchez, Gravit. Cosmol. \textbf{25}, (2019) 91–102 .}
\bibitem {mann}\emph{ E. Martin-Martinez, I.Fuentes, R.B. Mann, Phys.Rev.Lett. \textbf{107}, (2011) 131301.}
\bibitem {berry} \emph{M. V. Berry, Proc. Roy. Soc. A \textbf{392}, 45 (1984).}
\bibitem{berry1} \emph{M. V. Berry, J. of Phys. A: Math. and Gen., \textbf{18} (1985) 15-27.}
\bibitem {aranov}\emph{Y. Aharonov, J. Anandan, Phys. Rev. Lett. \textbf{58}, (1987) 1593
.}
\bibitem {tomita} \emph{ A. Tomita and R. Chiao, Phys. Rev.
Lett. \textbf{57}, (1986) 937;\\ D. Suter, G. C. Chingas, R. A. Harris and A. Pines, Mol. Phys. \textbf{61}, (1987) 1327;\\ D. Suter,
K. T. Mueller and A. Pines, Phys. Rev. Lett. \textbf{60}, 1218.
(1988);\\ J. A. Jones, V. Vedral, A. Ekert and G. Castagnoli, Nature \textbf{403}, (2000) 869.}
\bibitem{mina}\emph{S. K. Min, A. Abedi, K. S. Kim, E. K. U. Gross, Phys. Rev. Lett. \textbf{113}, (2014) 263004 .}
\bibitem{fuzi}\emph{K. Fuzikawa, Foundations of Quantum Mechanics in the Light of New Technology, pp. 290-293 (2006).}
\bibitem{hall}\emph{D. Bannerjee, P. Bandhyopadhyay, Mod. Phys. Lett.B, \textbf{8}, (1994) 26.}
\bibitem{zhang} \emph{Y. Zhang, Y.W. Tan, H. L. Stormer, P. Kim, Nature, Vol. 438, 201–204 (2005).}
\bibitem{Deri}\emph{A. Deriglazov, B. F. Rizzuti, American J. Phys. \textbf{79}, (2011) 882.}
\bibitem{dir}\emph{P. A. M. Dirac, Lectures on quantum mechanics, Vol. 2, Belfer Graduate School of
Science Monographs Series, Yeshiva University, New York, 1964.\\
A. Hanson, T. Regge and C. Teitelboim, Constrained Hamiltonian systems, Accademia
Nazionale dei Lincei (Rome, 1976).}
\bibitem{basu}\emph{ P. Basu, B. Chakraborty and F. G. Scholtz, J. Phys. A \textbf{44}, 285204 (2011).}
\bibitem{mehta}\emph{C. L. Mehta, E.C.G Sudarshan, Phys. Lett., \textbf{22} (1966) 5 .}
\bibitem{caru}\emph{P. Carruthers, M. M. Nieto, American J. Phys. \textbf{33}, 537 (1965).}
\bibitem{luk}\emph{J. Lukierski, P.C. Stichel, W.J. Zakrzewski, Ann. Physics \textbf{306} (2003) 78.}
\bibitem{sg}\emph{S. Gangopadhyay, F. G. Scholtz, J. Phys. A: Math. Theor.
\textbf{47} (2014) 235301.}
\bibitem{pn}\emph{P. Nandi, S. Sahu, S. K. Pal , Nucl. Phys. B \textbf{971} (2021) 115511 .}
\bibitem{biswa}\emph{F.G. Scholtz, B. Chakraborty, J. Govaerts, S. Vaidya,J.Phys.A \textbf{40} (2007) 14581-14592 .}
\bibitem{pnskp}\emph{S. Biswas, P. Nandi, B. Chakraborty, Phys.Rev.A, \textbf{102} (2020) 2, 022231.}
\bibitem{gauba}\emph{F.G. Scholtz, L. Gouba, A. Hafver, C.M. Rohwer, J.Phys.A, \textbf{42} (2009) 175303.}
\bibitem{chaoba}Y. C. Chaoba, B. Chakraborty, K. Kumar, F. G. Scholtz, Int. J. Geom. Meth. Mod. Phys. \textbf{15} (2018) 12, 1850204.
\bibitem{anwe}\emph{A. Chakraborty, B. Chakraborty, Int. J. Geom. Meth. Mod. Phys. \textbf{17} (2020) 06, 2050089 .}
\bibitem{povm}\emph{J.A. Bergou, Jnl. Phys. Conf. Series \textbf{84} (2007) 012001.}	
\bibitem{bal1}\emph{A.P. Balachandran, A. Ibort, G. Marmo, M. Martone, Phys.Rev.D, \textbf{81} (2010) 085017.\\
A.P. Balachandran, A. Ibort, G. Marmo, M. Martone, SIGMA \textbf{6} (2010) 052.}
\bibitem{sg2}\emph{S. Ghosh, Phys. Lett. B, \textbf{601} (2004) 93–98.}
\bibitem{bdr}\emph{ B. Dutta-Roy, G. Ghosh, J. of Phys. A: Math. and Gen. , \textbf{26} (1993), 1875-1879.} 
\bibitem{dasgupta}\emph{A. Dasgupta,  Am. J.  Phys. \textbf{64}, 1422 (1996).}
\bibitem{cast}\emph{E. Castellanos, J. I. Rivas, V. Dominguez-Rocha, EPL \textbf{106},(2014) 60005.}
\bibitem{bek}\emph{J. D. Bekenstein, Phys. Rev. D \textbf{86},(2012) 124040.}
\bibitem{anan}\emph{J Anandan, Y. Aharonov, Phys. Rev. Lett., \textbf{65}(1990) 14.} 
\bibitem{skb}\emph{S. K. Bose, B. Dutta-Roy, Phys. Rev. A, \textbf{43} (1991) 3217-3220.}
\bibitem{xu} \emph{J. B. Xu, X. C. Gao, Physica Scripta, \textbf{54} (1996) 137-139.}
\bibitem{nair}\emph{V.P. Nair, A.P. Polychronakos, Phys. Lett. B \textbf{505} (2001) 1-4, p.267-274.}
\bibitem{mohr}\emph{C. B. O. Mohr, Aust. J. Phys. , \textbf{10}, 01 (1957).}
\end{thebibliography}
\end{document}